\documentclass[aps,prb,twocolumn,showpacs,amsmath,amssymb,floats,showkeys,preprintnumbers,superscriptaddress]{revtex4-1}
\usepackage[dvips]{graphicx}
\usepackage{color}
\usepackage{float}
\usepackage{epsfig}
\usepackage{epstopdf}
\usepackage{ulem}
\usepackage{multirow}
\usepackage{dcolumn}
\usepackage{hyperref}
\usepackage[usenames,dvipsnames,svgnames,table]{xcolor}

\begin{document}
 
\title{Crystal  structure  search  and  electronic properties  of  alkali  doped
phenanthrene and picene}

\author{S.~Shahab Naghavi}
\affiliation{International  School for  Advanced  Studies  (SISSA), and  CNR-IOM
Democritos National Simulation Center, Via Bonomea 265, I-34136 Trieste, Italy }

\author{Erio Tosatti}
\affiliation{International  School for  Advanced  Studies  (SISSA), and  CNR-IOM
Democritos National Simulation Center, Via Bonomea 265,
I-34136  Trieste,  Italy  } 

\affiliation{International  Centre   for  Theoretical  Physics   (ICTP),  Strada
Costiera 11, I-34151 Trieste, Italy }

%\email[]{E-mail: tosatti@sissa.it }
%%%%%%%%%%%%%%%%%%%%%%%%%%%%%%%%%%%%%%%%%%%%%%%%%%%%%%%%%%%%%%%%%%%%%%%%%%%%%%%%%
\date{\today}
%%%%%%%%%%%%%%%%%%%%%%%%%%%%%%%%%%%%%%%%%%%%%%%%%%%%%%%%%%%%%%%%%%%%%%%%%%%%%%%%%
\begin{abstract} 
Alkali  doped   aromatic  compounds   have  shown   evidence  of   metallic  and
superconducting phases whose  precise nature is still  mysterious.  In potassium
and rubidium  doped phenanthrene,  superconducting temperatures around 5\,K have
been detected, but such basic  elements as the stoichiometry, crystal structure,
and electronic  bands are  still speculative.   We seek  to predict  the crystal
structure of M$_{3}$-phenanthrene (M\,=\,K,\,Rb)  using {\sl ab-initio} evolutionary
simulation in conjunction  with density functional theory (DFT),  and find metal
but  also insulator  phases  with distinct  structures.   The original  $P2_{1}$
herringbone structure of  the pristine molecular crystal  is generally abandoned
in favor  of different  packing and  chemical motifs.   The metallic  phases are
frankly ionic with three electrons acquired by each molecule. In the nonmagnetic
insulating phases the alkalis coalesce reducing the donated charge from three to
two per phenanthrene  molecule. A similar search for K$_3$-picene  yields an old
and a  new structure, with  unlike potassium positions and  different electronic
bands, but  both metallic retaining  the face-to-edge herringbone  structure and
the $P2_{1}$ symmetry  of pristine picene.  Both the new  K$_{3}$-picene and the
best metallic  M$_{3}$-phenanthrene are further  found to undergo  a spontaneous
transition from metal  to antiferromagnetic insulator when  spin polarization is
allowed, a  transition which is not  necessarily real, but which  underlines the
necessity to  include correlations beyond  DFT. Features of the  metallic phases
that may be relevant to phonon-driven superconductivity are underlined.
\end{abstract}
%%%%%%%%%%%%%%%%%%%%%%%%%%%%%%%%%%%%%%%%%%%%%%%%%%%%%%%%%%%%%%%%%%%%%%%%%%%%%%%%%
%\pacs{61.50.Ah,61.66.Hq,74.70.Kn}
\maketitle
%%%%%%%%%%%%%%%%%%%%%%%%%%%%%%%%%%%%%%%%%%%%%%%%%%%%%%%%%%%%%%%%%%%%%%%%%%%%%%%%%
\section{introduction}
\label{SEC:INTRO}  

The  field of  organic  superconductors  has received  recent  impulse with  the
discovery       of        superconductivity       in        electron       doped
picene,~\cite{mitsuhashi10,artioli14} promptly  followed by  a broader  class of
electron-doped   polycyclic   aromatic   hydrocarbons   (PAHs).    Among   them,
electron-doped                                coronene,~\cite{kubozono11,kato11}
1,2:8,9-dibenzopentacene,~\cite{xue12}  and phenanthrene~\cite{wang11,wang11a,
wang12} have been reported.  To this date, the
metallic  and superconducting  behavior of  many  of these  exciting systems  is
proving difficult to reproduce~\cite{mahns12,caputo12} and there is considerable
uncertainty,  ranging   from  stoichiometry,  mechanism,  and   precise  crystal
structure.  To  clarify the  situation, simultaneous  efforts of  experiment and
theory are called  for. The lack of knowledge of  stoichiometry and of structure
poses a starting dilemma to  theory, since calculations, either first principles
or modeling,  obviously depend on both  elements.  Most theoretical work  so far
has assumed the nominal three-electron stoichiometry that seemed closest to that
of the  superconducting cases, assuming  or optimizing crystal  structures where
six alkali-metal atoms fill a bimolecular unit cell of the same herringbone type
as the  pristine molecular insulator.~\cite{andres11} Based  on these reasonable
crystal structures, several superconductivity mechanisms have been discussed for
electron-doped PAHs.~\cite{subedi11,casula11,casula12,kato11,kosugi11,kosugi09,giovannetti11, Ruff013, Kim013,
naghavi14}  However, even  modest changes  in the  orientation and  deformation of  the PAHs
molecules and in  the relative position and distance of  alkali metal atoms with
respect to them and  among themselves can alter the bonding  nature of the doped
system and  its electronic structure,~\cite{gordon010} making  it somewhat risky
to  build  on  such  insecure  foundations.  Even  assuming  without  proof  the
M$_{3}$-PAHs  stoichiometry,  a structural  search  is  necessary.  The  crystal
structure searches so far appeared to preserve the pristine herringbone stacking
of PAH molecules,  limiting the freedom to the alkali  atom coordinates and thus
exploring only a very small portion of an extensive and complicated phase space.
Relaxing this assumption is important, especially as we shall see for small size
PAHs.

Predicting crystal  structures for  organic molecular  crystals is  not trivial.
Intermolecular forces are correlation-dominated and  thus hard to determine with
confidence by means of standard {\sl ab-initio} density-functional theory (DFT),
where correlations are  not automatically dealt with. The  problem is aggravated
by the  weakness of  these forces, which  gives rise to  near degeneracies  of a
confusing  variety   of  structures.   Connected  with   the  latter   is  frank
polymorphism, with many competing and coexisting crystalline structures with the
same  chemical   composition  but  different  symmetries   and  physico-chemical
properties,~\cite{Bern07-BOOK}  a possibility  likely to  be important  and that
should be seriously considered in alkali-doped  PAHs.  To explore the variety of
low energy structural  polymorphs an extensive search over  the energy landscape
of each  compound is essential.  To the best  of our knowledge,  search examples
from  which  criteria and  trends  could  be  learned  ahead of  hopeful  future
experimental  verifications are  not yet  satisfactorily developed  for electron
doped PAHs.

This  paper deals  with  crystal structure  prediction  for the  computationally
simpler              PAH             compounds              M$_{3}$-phenanthrene
(M$_{3}$-PA),~\cite{wang11,mitsuhashi10} and by comparison with the best studied
compound,  namely K$_{3}$-picene  (K$_{3}$-PC).~\cite{mitsuhashi10} To  keep the
search as  unbiased as possible  we resorted  to an evolutionary  algorithm (EA)
whose only required knowledge is chemical composition (see Sec.~\ref{SEC:METHOD}
for details).   It is based  on DFT supplemented either  by van der  Waals (vdW)
corrections,  which  account  approximately   for  some  long-range  correlation
effects, or  when more convenient  on a  hydrostatic pressure suitably  tuned to
yield  equivalent results  to vdW.   The small  unit cell  size and  the assumed
integrity of molecules  exclude by construction all  possibilities of structural
decomposition or polymerization. Despite  these limitations, the results reserve
interesting and hopefully relevant surprises.

%%%%%%%%%%%%%%%%%%%%%%%%%%%%%%%%%%%%%%%%%%%%%%%%%%%%%%%%%%%%%%%%%%%%%%%%%%%%%%%%%

\section{Computational Methods}
\label{SEC:METHOD}

Our search  of the stable and  low-energy metastable structures was  carried out
using   evolutionary    algorithm   (EA),   as   implemented    in   the   USPEX
code.~\cite{USPEX1,USPEX2,USPEX3,USPEX4,USPEX5M}    EA   was    implemented   in
conjunction  with  {\sl  ab-initio}   structure  relaxations  based  on  density
functional  theory  (DFT)  within  the  Perdew-Burke-Ernzerhof  (PBE)~\cite{PBE}
generalized gradient approximation as implemented in VASP ({\sl Vienna Ab-initio
Simulation Package}),  ~\cite{VASP1} employing  all-electron projector-augmented
plane  wave   (PAW)  method,~\cite{PAW1,PAW2}  as   well  as  on   {\sc  Quantum
ESPRESSO}.~\cite{pwscf,vdW-Stress-pwscf} Van  der Waals forces were  included in
two different implementations,  and separately compared to  a simple hydrostatic
pressure between 4 and 10\,Kbar whose  effects after tuning are found to be roughly
equivalent to vdW. The stronger chemical  bonding involving the alkali atoms, as
opposed to the  weaker and more delicate intermolecular bonding  of pristine PAH
crystals makes the use of DFT plus  van der Waals more reliable for alkali-doped
than  for  the   pristine  materials,  which  would   generally  require  better
treatments.~\cite{Hongo10}  Comparison  of  different  calculations  provides  a
picture  of the  general trends  which is  both instructive  and in  our opinion
trustworthy. The  energy cutoff for the  plane-wave basis was set  to 550\,eV to
ensure  full  convergence,  and  zero-point energies  were  not  included.   The
Brillouin  zone was  sampled by  Monkhorst-Pack  meshes with  the resolution  of
$2\pi\times 0.05$\,{\AA}$^{-1}$.   This resolution was decreased  to $2\pi\times
0.07$\,{\AA}$^{-1}$ to  limit the cost  of the calculation for  K$_{3}$-PC which
has a large  unit cell (78 atoms/cell).   As a check, we  recalculated the final
candidate structure with  a denser $k$-points sampling and  no noticeable change
of energy differences between structures emerged.

%%%%%%%%%%%%%%%%%%%%%%%%%%
%Fig. 1
\begin{figure}[htb!]
\centering
\includegraphics[width=0.99\linewidth]{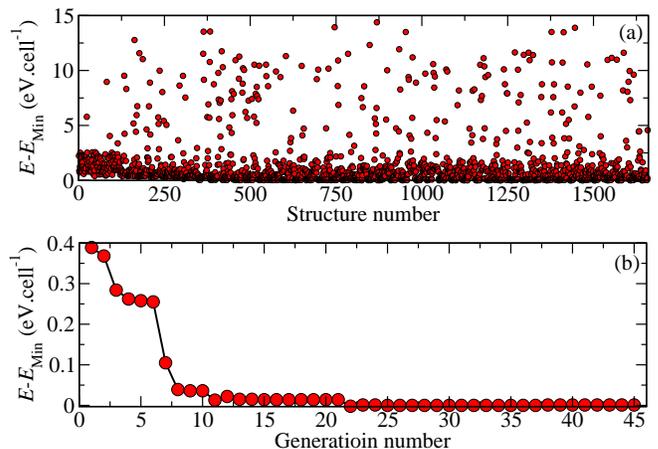}
\caption{(Color online)  (a) DFT-PBE total  energy per dimolecular cell  of over
1500  structures of  K$_{3}$C$_{14}$H$_{10}$ (K$_{3}$-PA  ) with  54 atoms/cell.
(b) Total energy  evolution with increasing generation number.}
\label{FIG:GENETIC}
\end{figure}
%%%%%%%%%%%%%%%%%%%%%%%%%%

A strictly trivalent stoichiometry M$_{3}$-PAH  was assumed, each unit cell used
in  the  optimization  procedure  containing  two  aromatic  molecules  and  six
identical alkali  atoms; the molecules  could deform  but not break.   The small
unit cell size, 54 atoms of M$_{3}$-PA (M\,=\,K, Rb) and 78 atoms of K$_{3}$-PC,
and the assumed integrity of molecules exclude by construction all possibilities
of  structural  decomposition  or  polymerization. For  the  structural  search,
conducted at     zero pressure    and  temperature, the initial  population (the
number of  structures   in    starting generation) to    densely   sample    the
configuration space.  For  M$_{3}$-PA we explored over  two thousands structures
within  about 60  generations  to  arrive at  the  converged  stable low  energy
structures.   Fig.~\ref{FIG:GENETIC} shows  the  energy  of approximately  1,500
structures (upper panel) calculated at 45 generations (lower panel). This coarse
search was done using straight DFT-PBE.

%%%%%%%%%%%%%%%%%%%%%%%%%%%%%%%%%%%%%%%%%%%%%%%%%%%%%%%%%%%%%%%%%%%%%%%%%%%%%%%%%
\section{Results }

In      previous      theoretical       studies      of      M$_{3}$-PA      and
K$_{3}$-PC~\cite{hansson06,kosugi09,
subedi11,casula11,kato11,kosugi11,giovannetti11,andres11,andres11pc,casula12,Ruff013, Kim013,naghavi14}
the doped  PAHs were  generally assumed  or in  some cases  found to  retain the
herringbone arrangements of  $P2_{1}$ symmetry similar to those  of the pristine
molecular  crystals  as  in   Fig.~\ref{FIG:STRUCT}.   That  assumption  is  not
unreasonable, but  it is unlikely  to be of  general validity, and  therefore we
released  it.   Especially for  the  smaller  size  PAH molecules  where  direct
intermolecular  forces  are  weaker  the   addition  of  six  alkali  atoms  per
bimolecular per cell  is a major perturbation that could  impact the herringbone
structure.   In our  EA  search for  M$_{3}$-PA (M\,=\,K,  Rb)  and K$_{3}$-PC,  the
molecular arrangement was  given freedom to change, and in  fact the herringbone
arrangement was found to survive in  K$_{3}$-PC but not in M$_{3}$-PA.  Below we
present results of  M$_{3}$-phenanthrene, (M\,=\,K, Rb) first, and for K$_{3}$-picene
second.   We  will present  structures,  their  energies, and  their  electronic
structures followed by a conclusive discussion.

%%%%%%%%%%%%%%%%%%%%%%%%%%%%%%%%%%%%%%%%%%%%%%%%%%%%%%%%%%%%%%%%%%%%%%%%%%%%%%%%%

\subsection{M$_{3}$-Phenanthrene (M\,=\,K, Rb)}

The behavior of  K$_{3}$-PA and of Rb$_{3}$-PA was found  to be totally similar,
once a volume  expansion of about 8\% is considered  between the two.  Therefore
we will  mostly discuss  K$_{3}$-PA here,  while all  conclusions are  valid for
Rb$_{3}$-PA too.

%%%%%%%%%%%%%%%%%%%%%%%%%%
%Fig. 2
\begin{figure}[htb!]
\centering
\includegraphics[width=0.99\linewidth]{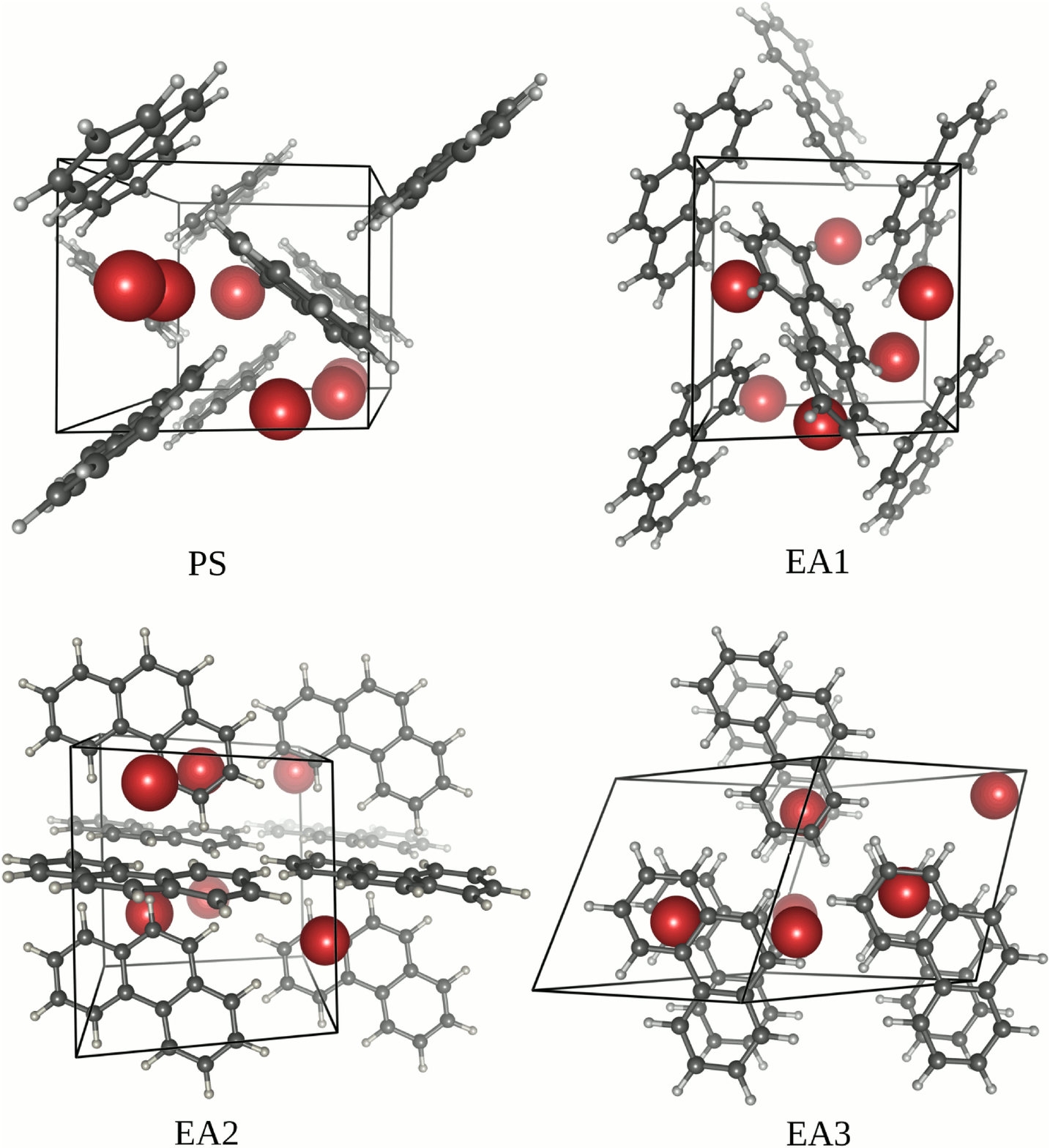}
\caption{(Color  online)  Crystal  structures   of  the  M$_{3}$-PA  low  energy
structures, obtained assuming two  PA molecules and six K or  Rb atoms per cell.
PS  is  the  previous  $P2_{1}$  symmetry herringbone  structure  of  de  Andres
et\,al,~\cite{andres11},  chosen as  reference.  Structure  EA1 (metallic)  is a
first evolution  of PS; EA2 (insulating)  and EA3 (metallic) are  the two lowest
energy structural polymorphs.  All have a  lower $P_{1}$ symmetry, but differ by
arrangement  and  shape  details  of  the phenanthrene  molecules,  and  by  the
positions of the alkalis.  Note  that approximately defined molecular planes are
nearly orthogonal in EA2, but have  turned parallel in EA3, which represents our
best candidate metallic (possibly superconducting) phase.}
\label{FIG:STRUCT}
\end{figure}
%%%%%%%%%%%%%%%%%%%%%%%%%

\begin{figure}[htb!]
\centering
\includegraphics[width=1.\linewidth]{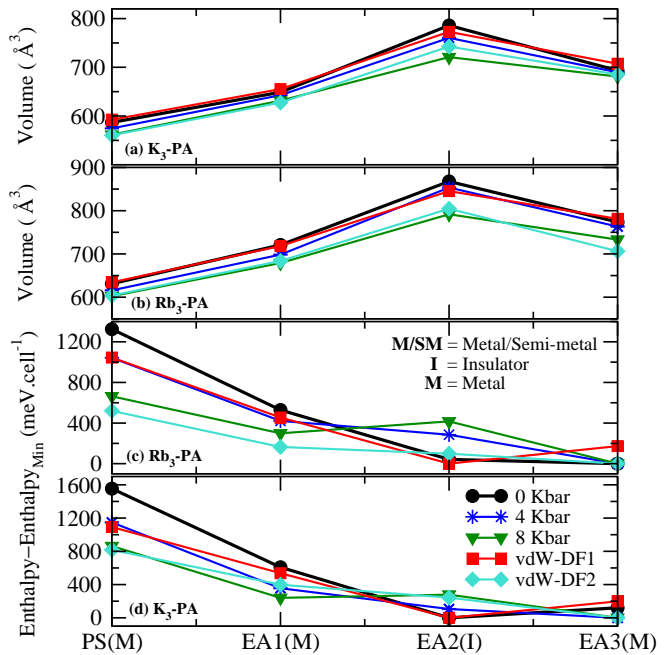}
\caption{(Color online) Cell  volume and enthalpy of the  most stable K$_{3}$-PA
and  Rb$_{3}$-PA candidate  structures.   After a  DFT-PBE  genetic search,  the
lowest energy  structures were refined  with vdW functionals or  with increasing
pressure.}
\label{FIG:VE-PA}
\end{figure}

We obtained the three lowest  energy structures EA1--3 of Fig.~\ref{FIG:STRUCT},
with lattice parameters listed in Table~\ref{TAB:PA}. Full crystal structure data
is given in the Supplementary Information in $cif$ format at the end of this manuscript.  
The three structures differ
in the  orientation (and distortion)  of the  phenanthrene molecules and  in the
position of the alkali atoms in the bimolecular cell.  Each of these structures,
generally with $P_{1}$ symmetry, represents a ``relevant structure", standing for
a  multiplicity  of  nearly  degenerate structures  differing,  e.g.,  by  small
displacements of alkali  atoms or small distortions of the  PA molecules.  Later
we  will provide  for the  best  metallic structure  EA3  an example  of such  a
slightly  distorted  structure   which  turns  the  crystal   from  metallic  to
semi-metallic.   Total  energies  and  structural parameters  are  presented  in
Fig.~\ref{FIG:VE-PA}, also illustrating the uncertainty due to the different vdW
functional  chosen, as  well as  a rough  equivalence of  some vdW  forces to  a
hydrostatic pressure.
%%%%%%%%%%%%%%%%%%%%%%%%%%
%Fig. 3
%%%%%%%%%%%%%%%%%%%%%%%%%%

\begin{figure*}[htp!]
\centering
\includegraphics[width=0.99	\linewidth]{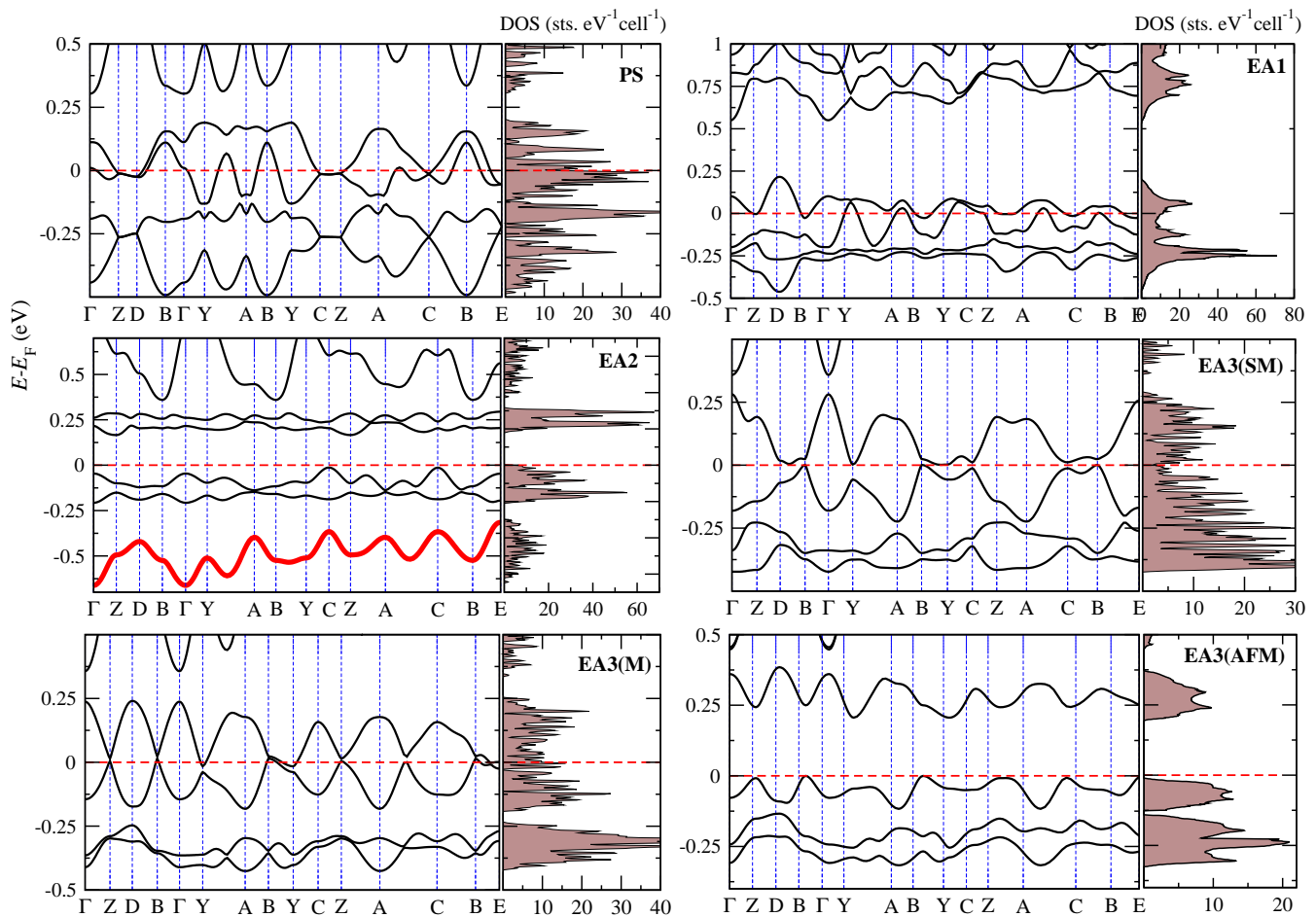}
\caption{(Color online) DFT-PBE electronic  band structure of various K$_{3}$-PA
crystal  structures.  Among  the  two best  structures  EA2 and  EA3,  EA2 is  a
nonmagnetic band insulator, where the  two uppermost filled bands have molecular
LUMO character,  the lowest unfilled  bands a  LUMO+1 character, and  the filled
band emphasized in  red is a shared  orbital within the six alkali  atoms in the
cell.  EA3 instead is the best  metallic structure with partly degenerate LUMO+1
half   filled  bands   at  Fermi   level,  not   unlike  those   calculated  for
La-PA.~\cite{naghavi14}. In analogy  to that case, the  near-degeneracy at Fermi
can be reduced or lifted by a slight intra-cell atomic displacement (here of the
K  atoms) which  takes  the system  to a  semi-metallic  state.  The  insulating
antiferromagnetic electronic structure obtained by allowing spin polarization in
metallic EA3 is also shown.}
\label{FIG:BAND-PA}
\end{figure*}

The first relevant structure EA1 is  a herringbone although a different one from
PS  ~\cite{andres11}  with  a  0.4--0.8\,eV/cell   lower  energy,  as  seen  in
Fig.~\ref{FIG:VE-PA}.  Molecules  in EA1  are T-shape  oriented and  are stacked
long edge-to-face, with an angle  of about 70$^{\circ}$.  The DFT-PBE electronic
structure calculation (Fig.~\ref{FIG:BAND-PA}) shows that EA1  is  metallic, but 
with much narrower bands and lower symmetry than PS.

%%%%%%%%%%%%%%%%%%%%%%%%%%
%Fig. 4
%%%%%%%%%%%%%%%%%%%%%%%%%%%%%%%%%%%%%%%%%%%%%%%%%%%%%%%%%%%%%%%%%%%%%%%%%%%%%%%%%

\begin{figure}[htb!]
\centering
\includegraphics[width=0.99\linewidth]{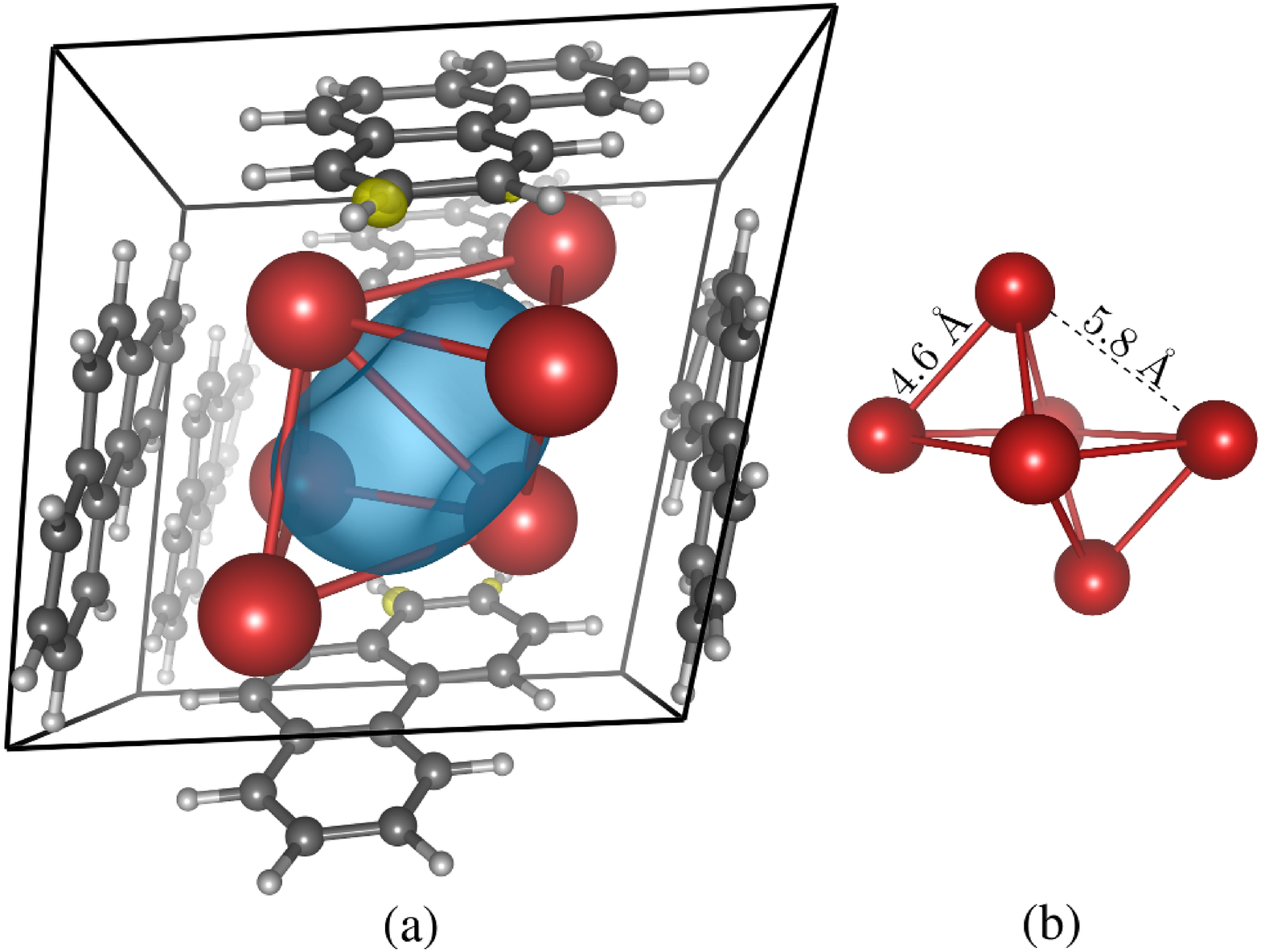}
\caption{(Color online)  Electron density  portrait $|\Psi(r)|^2$ of  the filled
orbital shown as a red band in Fig.~\ref{FIG:BAND-PA}  for the  EA2  nonmagnetic
insulating structure  of K$_{3}$-PA.  Coordinates  were chosen to  emphasize the
shared nature of this orbital between the six clustered K atoms.}
\label{FIG:BONDING}
\end{figure}

In  structure EA2  the  PA  molecules approximately  lie  on alternating  nearly
orthogonal planes and are relativel flat, but the six  alkali  atoms  are grouped
close  enough  together  to  form a sort of loose cluster, sitting between the molecules. 
The formation  energy  of M$_{3}$-PA compounds  relative  to
the  metallic alkali  metal plus  pristine phenanthrene is definitely negative,~\cite{Yan2014}
which rules out phase separation, which would be surprising here even if in an embryonic form. 
In order to understand what happens we  carried out an analysis  
of the bonding which is shown in Fig.\ref{FIG:BONDING}, and of the resulting electronic structure,
Fig.\ref{FIG:BAND-PA}.

%%%%%%%%%%%%%%%%%%%%%%%%%%%%%%%%%%%%%%%%%%%%%%%%%%%%%%%%%%%%%%%%%%%%%%%%%%%%%%%%%%%
% Fig.5

%%%%%%%%%%%%%%%%%%%%%%%%%%%%%%%%%%%%%%%%%%%%%%%%%%%%%%%%%%%%%%%%%%%%%%%%%%%%%%%%%

The outcome is  that in EA2  the M$_{6}$ alkali  cluster as  a whole
donates four of its six $s$-electrons to the LUMO  orbitals of the two PAH
molecules in the  cell, leaving  the LUMO+1  totally empty.  The two  remaining
alkali  $s$-electrons fill  up instead  a  ``superatomic" orbital  of the  alkali
cluster, emphasized in  Fig.\ref{FIG:BAND-PA}, with energy well  below the Fermi
energy and even below the LUMO band. This electronic configuration gives rise to
an  insulator  with  a  gap   of  0.2--0.4\,eV,  roughly  corresponding  to  the
LUMO-LUMO+1 energy  separation.  The EA2  structure is therefore not  a straight
phase separation in an embryonic stage but rather appears to represent, with all
the caution  and the provisos suggested  by assumptions made about  cell size and
about stoichiometry, a true polymorph of M$_{3}$-PA.  Its structure enhances the
crystal volume  with a  large intermolecular  distance between  the phenanthrene
molecules,  yet   the  mixed  ionic-covalent  bonding   makes  it  energetically
competitive. Of course, unusual as this structure is, it would need checking, e.g., 
by increasing the simulation cell size, for example by doubling it in the form (2x2x2).
However, that would call for a 2$^6$ times large computational effort which is 
presently beyond the reach of this work. Because of that, and since EA2 is not always 
the lowest energy structure  at least in presence of van der Waals forces,  and is not
even metallic, we do not pursue it further.

\begin{figure}[htb!]
\centering
\includegraphics[width=0.99\linewidth]{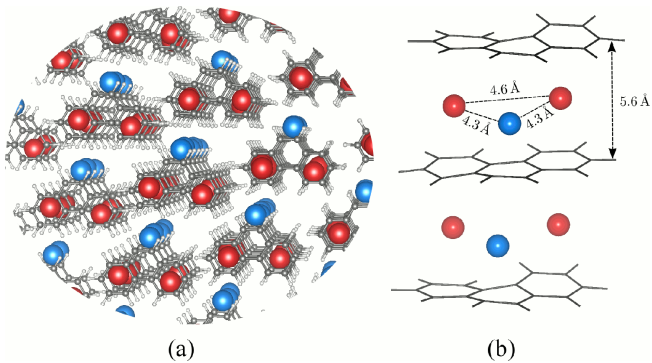}
\caption{(Color online) Top view of the EA3 structure of K$_{3}$-PA. K atoms are
aligned in  a slightly deformed  straight line with  a K$_{1}$--K$_{2}$--K$_{3}$
angle of 174$^{\circ}$.  An increased  zig-zag deformation reducing this angle to
162$^{\circ}$ transforms the  metal into a semi-metal. The  distance between the
phenanthrene  molecular centers  is about  5.5\,{\AA},  that between  the K  and
center  of the  nearest  benzene   ring of  PA  2.7\,{\AA}.   The PA  molecules,
although not strictly planar, approximately  lie on parallel planes, in contrast
with the herringbone structure of pristine phenanthrene and of structure EA1.}
\label{FIG:EA3}
\end{figure}

Structure  EA3 finally,  our  best  metallic structure,  is  characterized by  a
remarkably  different,  parallel  stacking   of  phenanthrene  molecular  layers
repeating  on top  of one  another (AAA)  with alkali  atoms spread  between the
layers   as    in   Fig.~\ref{FIG:STRUCT}.    The   electronic    structure   of
Fig.\ref{FIG:BAND-PA} shows half-filled  LUMO+1 bands of width  0.2\,eV that are
half-filled  and   somewhat  close  to  electron-hole   symmetrical,  with  near
degeneracies and a high density of states  near the Fermi level. As was the case
in    a    model   previously    derived    for    the   hypothetical    crystal
La-phenanthrene,~\cite{naghavi14} this  situation suggests a  potentially strong
Fr{\"o}hlich electron-phonon  coupling to  zone boundary  modes.  Among  them, a
distortion  that would  reduce  the  approximate equidistance  of  atoms and  of
molecules   in   the   cell,   removes   this  degeneracy   and   leads   to   a
semimetal.~\cite{naghavi14}  Modifying   the  structure   EA3  of   K$_3$-PA  by
statically  displacing  the alkali  atoms  out  of their  optimal  straight-line
arrangement  (Fig.~\ref{FIG:EA3})  for  example  to   form  an  angle  of  about
174$^{\circ}$ rather  than 180$^{\circ}$, the  band splitting and  the resulting
semimetallic behavior obtained  are shown in Fig.~\ref{FIG:BAND-PA}.   This is a
suggestive  element  in  view  of   a  possible  electron-phonon  mechanism  for
superconductivity of  electron-doped PAHs,  which is presently  under discussion
and is  being addressed elsewhere.~\cite{qin14} The  gap-opening distortion just
chosen  here  is  of  course  somewhat   arbitrary,  and  a  proper  phonon  and
electron-phonon calculation will be needed in order to pursue this line.

%%%%%%%%%%%%%%%%%%%%%%%%%%
%Fig. 6
%%%%%%%%%%%%%%%%%%%%%%%%%%

%%%%%%%%%%%%%%%%%%%%%%%%%%
\begin{table}[htb!]
\centering
\caption{Energy and  structural data  of the best  structures of  K$_{3}$-PA and
Rb$_{3}$-PA calculated with DFT- GGA at  zero pressure compared with the earlier
proposed  PS structure.~\cite{andres11}.   Energy  in  units of  meV.cell$^{-1}$
measured with  respect to that of  the best structure.  Volume  is in \AA$^{3}$.
Each of these  relevant structures generally allows  slightly different variants
depending on small extra alkali or molecular displacements.}
\begin{ruledtabular}
\begin{tabular}{cccccc}
           & PS(M) & EA1(M)  & EA2(I) &EA3(SM)& EA3(M) \\ \hline
\multicolumn{6}{c}{K$_{3}$-PA} \\
Sym.       & P21   & P21    &   P1    & P1    & P1     \\
Energy     & 1352  & 556    &   70    & 28    & 0      \\
Volume     & 586.9 & 648.8  &   786.0 & 693.4 & 725.7  \\
$a$        & 8.24  & 8.95   &   9.63  & 9.96  & 10.20  \\
$b$        & 7.05  & 8.32   &   9.89  & 8.92  & 9.12   \\
$c$        & 10.7  & 8.9    &   9.5   & 9.5   & 9.6    \\
$\alpha$   & 90.0  & 103.0  &   112.6 & 95.1  & 97.2   \\
$\beta$    & 108.8 & 90.0   &   90.7  & 114.2 & 107.8  \\
$\gamma$   & 90.0  & 90.0   &   100.7 & 64.9  & 116.2  \\
\multicolumn{6}{c}{} \\
\multicolumn{6}{c}{Rb$_{3}$-PA} \\
Sym.       & P21   & P21    &   P1    & P1    & P1     \\  
Energy     & 1554  & 608    &   0     & 129   & 120    \\  
Volume     & 631   & 721    &   868   & 765   & 774    \\  
$a$        & 8.74  & 9.16   &   9.87  & 10.32 & 10.23  \\  
$b$        & 7.10  & 8.64   &   10.12 & 9.13  & 9.18   \\  
$c$        & 10.78 & 9.26   &   9.64  & 9.82  & 9.79   \\  
$\alpha$   & 90.0  & 100.6  &   113.4 & 94.9  & 96.0   \\  
$\beta$    & 109.4 & 90.0   &   90.8  & 112.2 & 106.2  \\  
$\gamma$   & 90.0  & 90.0   &   99.9  & 63.7  & 115.0  \\  
\end{tabular}
\end{ruledtabular}
\label{TAB:PA}
\end{table}
%%%%%%%%%%%%%%%%%%%%%%%%%%

\subsection{K$_{3}$-Picene}

The genetic crystal structure optimization  just conducted for M$_{3}$-PA led to
identify  unexpected  structures,  where   the  pristine  herringbone  molecular
stacking is  abandoned.  In view of  that, it is worthwhile  conducting the same
search  for K$_{3}$-PC.   Repeating  for  better confidence  the  search with  a
variety of different  starting points and both with DFT-vdW  and with DFT-PBE at
10\,kbars which are  roughly equivalent, we invariably found  in K$_{3}$-picene a
face-to-edge  herringbone  packing  as  the   best  structure,  in  contrast  to
M$_{3}$-PA.   This supports  the assumptions  and the  results ,  anticipated by
previous authors, that the basic herringbone structural motif of pristine picene
is preserved.~\cite{andres11pc,Ruff013}. 

Previous  studies  of  acenes   emphasized  the
importance of  the balance between  electrostatic and van der  Wals interactions
for the  packing in  molecular crystals.~\cite{Kafer08,Janowski10}  As discussed
for  pentacene,~\cite{Kafer08}  co-planar   stacking  essentially  results  from
electrostatic  interaction  while  face-to-edge  herringbone  is  stabilized  by
dispersion and quadrupole forces.  The much larger size of picene implies larger
quadrupolar  forces,  which explains  the  large  stability of  herringbone,  as
opposed to the fragility against doping just shown for phenanthrene.

%%%%%%%%%%%%%%%%%%%%%%%%%%%%%%%%%
% Fig. 7
\begin{figure*}[htb!]
\centering
\includegraphics[width=0.99\linewidth]{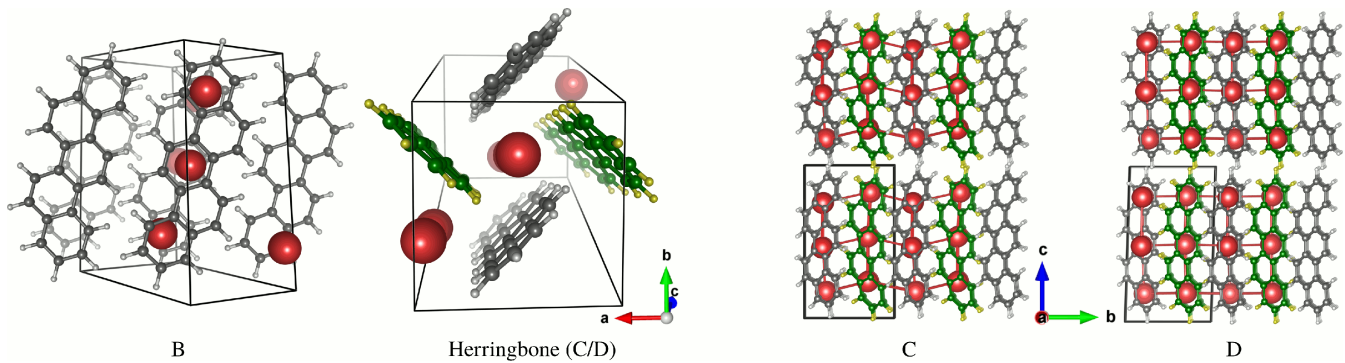}
\caption{(Color online) The three best structures B,C, D of K$_{3}$-picene.  All
structures  are herringbone  with  $P2_{1}$ symmetry  as  in pristine  molecular
picene.     D     basically    coincides    with    a     previously    proposed
structure\cite{andres11pc}, whereas  in C  alternating lines of  potassium atoms
are shifted relative  to one another, forming  a zig-zag network.  C  and D have
basically the same total energy.}
\label{FIG:K3PC-STR}
\end{figure*}
%%%%%%%%%%%%%%%%%%%%%%%%%%%%%%%%%%%%%%%%%%%%%%%%%%%%%%%%%%%%%%%%%%%%%%%%%%%%%%%%%

%%%%%%%%%%%%%%%%%%%%%%%%%%%%%%%%%%%%%%%%%%%%%%%%%%%%%%%%%%%%%%%%%%%%%%%%%%%%%%%%%
% Fig. 8
\begin{figure}[htb!]
\centering
\includegraphics[width=0.99\linewidth]{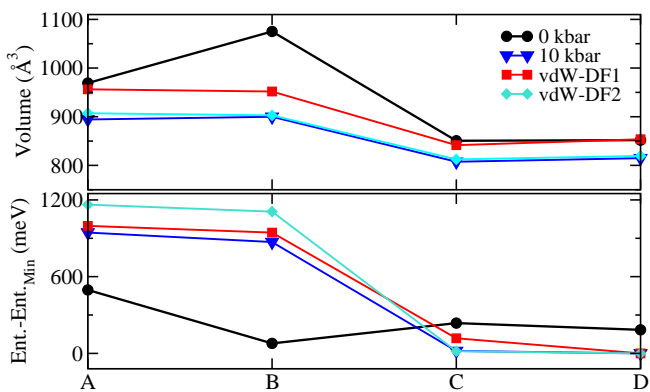}
\caption{(Color online) Volume (a) and  enthalpy (b) of four relevant structures
A--D  of K$_{3}$-PC  using  different functionals  and  at different  pressures.
Structure    D    is   closest    to    earlier    established   structure    in
Ref.~\cite{andres11pc} The two best structures C  and D are nearly degenerate in
energy and  differ by the  fractional shift of one  alkali line relative  to the
other line in the  unit cell. Note that here vdW  corrections are more important
than  in M$_{3}$-PA,  on account  of the  larger molecular  size. A  hydrostatic
pressure of about 10\,kbars (blue line) has here a similar effect to vdW-DF (cyan
and red lines).}

\label{FIG:K3PC-ENE}
\end{figure}
%%%%%%%%%%%%%%%%%%%%%%%%%%%%%%%%%%%%%%%%%%%%%%%%%%%%%%%%%%%%%%%%%%%%%%%%%%%%%%%%%

%%%%%%%%%%%%%%%%%%%%%%%%%%%%%%%%%%%%%%%%%%%%%%%%%%%%%%%%%%%%%%%%%%%%%%%%%%%%%%%%%
%Fig. 9
\begin{figure*}[htb!]
\centering
\includegraphics[width=0.99\linewidth]{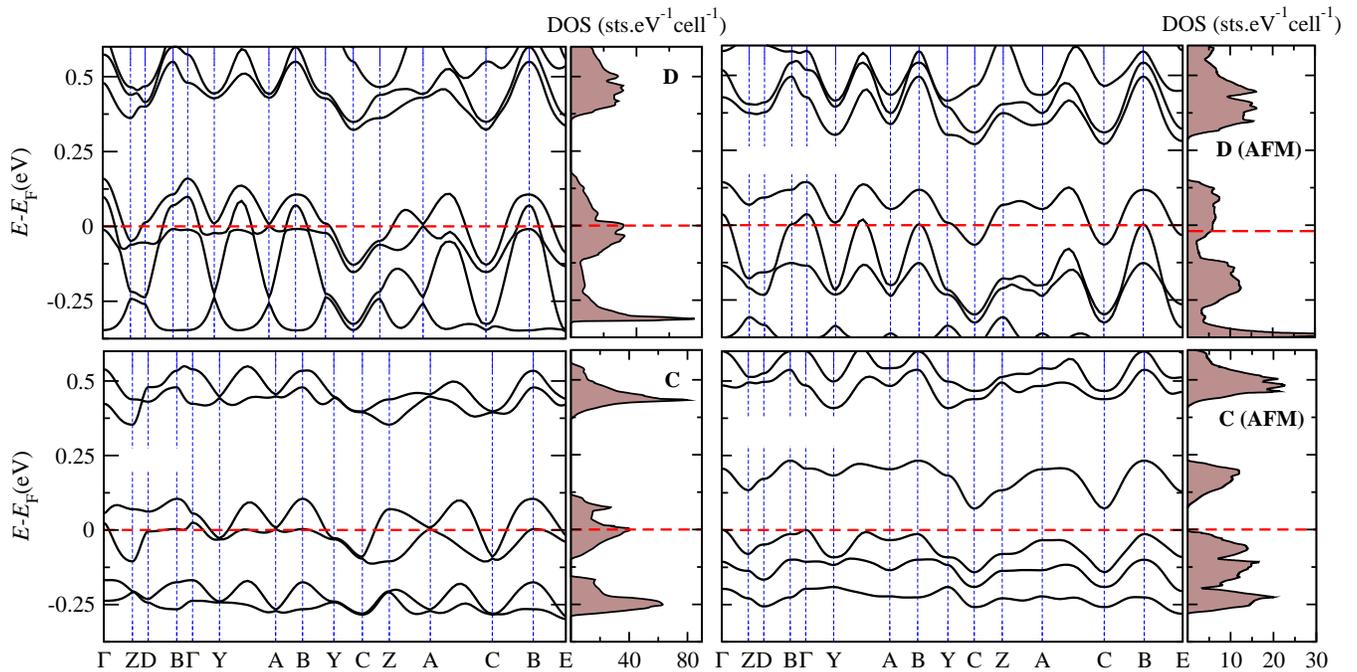}
\caption{(Color  online) Electronic  band structure  of  structures C  and D  of
K$_{3}$-PC  in  the non-magnetic  (left)  and  antiferromagnetic state  (right).
Structure C  has narrower  bands, and  gains 40\,meV  per dimolecular  cell upon
antiferromagnetic ordering,  while in the  broader-band crystal structure  D the
gain is about  20\,meV. The energy lowering in  the antiferromagnetic insulators
signals the importance of correlations, more than a real magnetic instability of
the metallic state.~\cite{qin14}}
\label{FIG:K3PC-BAND}
\end{figure*}
%%%%%%%%%%%%%%%%%%%%%%%%%%%%%%%%%%%%%%%%%%%%%%%%%%%%%%%%%%%%%%%%%%%%%%%%%%%%%%%%%

Both C  and D  structures, 
that are energetically the best in presence of van der Waals forces, 
are face-to-edge  herringbone with  $P2_{1}$ symmetry (Fig.\ref{FIG:K3PC-STR} and Fig.\ref{FIG:K3PC-ENE}).
Similarly to the previously optimized K$_{3}$-PC-structure~\cite{andres11pc} the
potassium atoms  form ordered  chains along  the $c$-axis, lying  in front  of the
aromatic ring (Fig.\ref{FIG:K3PC-STR}). In  structure D the chains are uniformly  stacked, whereas in the
ordering is  zigzag shifted along  the $b$-axis, so  that the potassium  atoms are
located in front of the C--C bond instead of the aromatic ring center.

In spite  of the similarity  between the  two structures, their  electronic band
structures  are quite  different.  As  seen in  Fig.~\ref{FIG:K3PC-BAND}, C  has
narrow  bandwidths  (the  strongest hopping  parameter in  the  LUMO+1 band  is
0.02\,eV) and LUMO, and LUMO+1 are narrow  and well separated. A high density of
states at Fermi energy together with the narrow bandwidth lead to instability in
the  system, prone  to  magnetism or  to  a structural  distortion,  or else  to
superconductivity.  We find that antiferromagnetic ordering lowers the energy of
C by 40\,meV/bimolecular~cell,  whereas the antiferromagnetic  total energy gain
in  D  is  only  20\,meV/bimolecular~cell,
a smaller value due to residual band overlap and metallicity.

We  stress here  that this  kind antiferromagnetic  instability of  the metallic
phases,  well   known  and  quite   general  for  these  narrow   band  systems,
\cite{giovannetti11}  must be  considered with  caution. It  is surely  a signal
warning  that correlations  are  very strong,  but  it needs  not  imply a  true
instability of  the metallic  state. As  is independently  being shown  e.g., in
Ref~\cite{qin14} the inclusion of correlations through a Gutzwiller treatment of
a  Hubbard  $U$ can  upset  the  energetic balance  between  the  metal and  the
antiferromagnetic  insulator,  actually  favoring a  nonmagnetic,  metallic  and
potentially  superconducting state  at  least for  small  $U$.  This  background
knowledge allows  the provisional neglect  of the magnetic instabilities,  and a
serious  consideration of  crystal  structures  C and  D  as genuine  candidates
describing the metallic phase of K$_{3}$-PC.

%%%%%%%%%%%%%%%%%%%%%%%%%%%%%%%%%%%%%%%%%%%%%%%%%%%%%%%%%%%%%%%%%%%%%%%%%%%%%%%%%
\section{Conclusions}  

Electron doped PAHs  encompass a larger variety of  crystal structure polymorphs
than  imagined  so far.  We  systematically  studied  the crystal  structure  of
M$_{3}$-PA (M  = Rb and  K), an  interesting system where  superconductivity has
been  reported,  and  of  K$_{3}$-PC,   the  most  studied  electron  doped  and
superconducting PAH.   By means  of an evolutionary  algorithm we  identified in
K$_{3}$-PA  and Rb$_{3}$-PA  at least  two possible  polymorphs with  distinctly
different structures that are competing for the ground state.  One of them (EA2)
is  a  band  insulator, the  other  (EA3)  is  metallic,  and thus  a  candidate
superconductor.   Structurally,  the  pristine herringbone  structural  symmetry
appears systematically broken by the alkali doping.

The scenario is different in K$_{3}$-PC,  where the larger size molecule allows
to the  herringbone structure  and $P2_{1}$  symmetry to  survive in  the stable
structure. Here  too we still  find two essentially degenerate  polymorphs, both
metallic, whose only difference is the arrangement of the potassium atoms.

Within DFT-PBE all metallic  structures are unstable against antiferromagnetism,
an  instability  which  leads  them  to an  insulating  state  in  pin-polarized
DFT.  While that  signals  the  importance of  correlations  in the  nonmagnetic
metallic states, it does not imply their irrelevance or non-existence, as recent
Gutzwiller calculations have suggested.~\cite{qin14}

Should  the  three-electron  stoichiometry  which we  assumed  beconfirmed,  the
present  results  would also  be  relevant  to  the superconductivity  of  these
compounds.  The electronic  structures of most metallic phases  we found, either
M$_{3}$-PA or  K$_{3}$-PC, share  a common element,  namely the  near degeneracy
among the  two LUMO+1  half-filled conduction  bands, quite  close to  the Fermi
level.   That  degeneracy may,  as  was  noted  in  an earlier  calculation  for
hypothetical La-PA, ~\cite{naghavi14} and as  is pointed out above for structure
EA3 of K$_{3}$-PA be  very efficiently split by a zone  boundary phonon. That is
in broad analogy  with MgB$_{2}$,  potentially interesting as  a driving element
of phonon-driven  superconductivity.  Our  recent study of  a ``Hubbard-Frohlich"
model derived  by idealizing this idea  is showing a $s_{  \pm}$ superconducting
phase     which    is     especially     robust    against     electron-electron
repulsion.~\cite{qin14}.

Comparison of  our theoretical crystal structures  with experiments, problematic
at the moment because well  defined single phase stoichiometrically defined data
are still being sought, will be  very interesting suggesting a close cooperation
with future  structural determinations.  We  underline the fact  that structural
work  and  chemical  understanding  represent an  absolutely  vital  and  urgent
necessity for the development and credibility of this field.  The present theory
work is a basic contribution to  this development, that will hopefully encourage
future experimental work.

%%%%%%%%%%%%%%%%%%%%%%%%%%%%%%%%%%%%%%%%%%%%%%%%%%%%%%%%%%%%%%%%%%%%%%%%%%%%%%%%%
\begin{acknowledgments}  
This  work was  supported by  the European  Union FP7-NMP-2011-EU-Japan  project
LEMSUPER and  in part by PRIN/COFIN  Contract PRIN-2010LLKJBX\_004 as well  as a
collaboration with  Argonne National  Laboratories.  We  acknowledge discussions
with LEMSUPER members, with our SISSA  group members M. Capone, M. Fabrizio, and
T.  Qin.  A  CINECA  award  2013 of  high  performance  computing resources  was
instrumental to the realization of this work.
\end{acknowledgments}
%%%%%%%%%%%%%%%%%%%%%%%%%%%%%%%%%%%%%%%%%%%%%%%%%%%%%%%%%%%%%%%%%%%%%%%%%%%%%%%%%
%%\bibliography{database_in_org.bib}

\pagebreak
\section*{Supplementary: \\ Crystallographic Information File (CIF)}

\subsection{M$_{3}$-phenanthrene}

\subsubsection{Structure \textbf{EA1}}
\begin{tiny}
\begin{verbatim}
#======================================================================

# CRYSTAL DATA

#----------------------------------------------------------------------

STRUCTURE_EA1


_pd_phase_name                         'EA128   8.936  8.282  8.918 102.934 89'
_cell_length_a                         8.94969
_cell_length_b                         8.31950
_cell_length_c                         8.94544
_cell_angle_alpha                      103.03106
_cell_angle_beta                       89.53737
_cell_angle_gamma                      89.61511
_symmetry_space_group_name_H-M         'P 1'
_symmetry_Int_Tables_number            1

loop_
_symmetry_equiv_pos_as_xyz
   'x, y, z'

loop_
   _atom_site_label
   _atom_site_occupancy
   _atom_site_fract_x
   _atom_site_fract_y
   _atom_site_fract_z
   _atom_site_adp_type
   _atom_site_B_iso_or_equiv
   _atom_site_type_symbol
   C1         1.0     0.236961      0.768231      0.097236     Biso  1.000000 C
   C2         1.0     0.195296      0.631071      0.156179     Biso  1.000000 C
   C3         1.0     0.149564      0.813722      0.981541     Biso  1.000000 C
   C4         1.0     0.063940      0.541797      0.099717     Biso  1.000000 C
   C5         1.0     0.012977      0.729514      0.925556     Biso  1.000000 C
   C6         1.0     0.970017      0.585593      0.987782     Biso  1.000000 C
   C7         1.0     0.921665      0.775133      0.816305     Biso  1.000000 C
   C8         1.0     0.561874      0.319951      0.831266     Biso  1.000000 C
   C9         1.0     0.654613      0.266666      0.936170     Biso  1.000000 C
   C10        1.0     0.601150      0.463035      0.776453     Biso  1.000000 C
   C11        1.0     0.787108      0.356064      0.988246     Biso  1.000000 C
   C12        1.0     0.740083      0.550975      0.820339     Biso  1.000000 C
   C13        1.0     0.833489      0.496517      0.935155     Biso  1.000000 C
   C14        1.0     0.787643      0.685189      0.762353     Biso  1.000000 C
   C15        1.0     0.746751      0.849955      0.274673     Biso  1.000000 C
   C16        1.0     0.695812      0.981149      0.211726     Biso  1.000000 C
   C17        1.0     0.662118      0.802331      0.392553     Biso  1.000000 C
   C18        1.0     0.559853      0.064903      0.268233     Biso  1.000000 C
   C19        1.0     0.519386      0.878206      0.445960     Biso  1.000000 C
   C20        1.0     0.470964      0.021866      0.385538     Biso  1.000000 C
   C21        1.0     0.426506      0.826122      0.551618     Biso  1.000000 C
   C22        1.0     0.061242      0.287455      0.556955     Biso  1.000000 C
   C23        1.0     0.158259      0.349585      0.460330     Biso  1.000000 C
   C24        1.0     0.099994      0.138562      0.602909     Biso  1.000000 C
   C25        1.0     0.293068      0.262513      0.405977     Biso  1.000000 C
   C26        1.0     0.240709      0.051779      0.556750     Biso  1.000000 C
   C27        1.0     0.336075      0.112868      0.446332     Biso  1.000000 C
   C28        1.0     0.289215      0.912825      0.607573     Biso  1.000000 C
   H1         1.0     0.328881      0.847007      0.150002     Biso  1.000000 H
   H2         1.0     0.464008      0.246700      0.786374     Biso  1.000000 H
   H3         1.0     0.256597      0.599057      0.250906     Biso  1.000000 H
   H4         1.0     0.180704      0.921418      0.936795     Biso  1.000000 H
   H5         1.0     0.946413      0.889846      0.780423     Biso  1.000000 H
   H6         1.0     0.035213      0.435953      0.147467     Biso  1.000000 H
   H7         1.0     0.622168      0.166956      0.990034     Biso  1.000000 H
   H8         1.0     0.540515      0.490738      0.678956     Biso  1.000000 H
   H9         1.0     0.721694      0.723199      0.673422     Biso  1.000000 H
   H10        1.0     0.854418      0.312302      0.072555     Biso  1.000000 H
   H11        1.0     0.842408      0.775102      0.221199     Biso  1.000000 H
   H12        1.0     0.963190      0.360829      0.606205     Biso  1.000000 H
   H13        1.0     0.754518      0.011553      0.114084     Biso  1.000000 H
   H14        1.0     0.697161      0.694501      0.435412     Biso  1.000000 H
   H15        1.0     0.455076      0.711419      0.586250     Biso  1.000000 H
   H16        1.0     0.523525      0.167037      0.218488     Biso  1.000000 H
   H17        1.0     0.130211      0.459584      0.419048     Biso  1.000000 H
   H18        1.0     0.036290      0.104707      0.696416     Biso  1.000000 H
   H19        1.0     0.222146      0.868650      0.693039     Biso  1.000000 H
   H20        1.0     0.366624      0.316211      0.332679     Biso  1.000000 H
   K1         1.0     0.042110      0.074333      0.232624     Biso  1.000000 K
   K2         1.0     0.735952      0.152823      0.566958     Biso  1.000000 K
   K3         1.0     0.242341      0.464780      0.801586     Biso  1.000000 K
   K4         1.0     0.541464      0.570971      0.137380     Biso  1.000000 K
   K5         1.0     0.014265      0.747870      0.476931     Biso  1.000000 K
   K6         1.0     0.515592      0.879690      0.883883     Biso  1.000000 K
\end{verbatim}
\end{tiny}
\vspace{2cm}

\subsubsection{Structure \textbf{EA2}}
\begin{tiny}
\begin{verbatim}
#======================================================================

# CRYSTAL DATA

#----------------------------------------------------------------------

STRUCTURE_EA2


_pd_phase_name                         'EA376   9.874 10.123  9.642 113.457 90'
_cell_length_a                         9.43072
_cell_length_b                         9.88197
_cell_length_c                         9.30359
_cell_angle_alpha                      112.95712
_cell_angle_beta                       91.82224
_cell_angle_gamma                      98.41454
_symmetry_space_group_name_H-M         'P 1'
_symmetry_Int_Tables_number            1

loop_
_symmetry_equiv_pos_as_xyz
   'x, y, z'

loop_
   _atom_site_label
   _atom_site_occupancy
   _atom_site_fract_x
   _atom_site_fract_y
   _atom_site_fract_z
   _atom_site_adp_type
   _atom_site_B_iso_or_equiv
   _atom_site_type_symbol
   C1         1.0     0.593349      0.322232      0.633825     Biso  1.000000 C
   C2         1.0     0.619814      0.179385      0.610941     Biso  1.000000 C
   C3         1.0     0.576455      0.362860      0.507546     Biso  1.000000 C
   C4         1.0     0.619213      0.073577      0.453311     Biso  1.000000 C
   C5         1.0     0.593044      0.263012      0.347730     Biso  1.000000 C
   C6         1.0     0.600691      0.108385      0.322594     Biso  1.000000 C
   C7         1.0     0.598036      0.306536      0.222096     Biso  1.000000 C
   C8         1.0     0.584097      0.784330      0.844042     Biso  1.000000 C
   C9         1.0     0.591501      0.736991      0.967879     Biso  1.000000 C
   C10        1.0     0.577304      0.932213      0.872624     Biso  1.000000 C
   C11        1.0     0.590464      0.844729      0.122744     Biso  1.000000 C
   C12        1.0     0.587054      0.047263      0.030257     Biso  1.000000 C
   C13        1.0     0.592111      0.996603      0.158627     Biso  1.000000 C
   C14        1.0     0.593137      0.198546      0.062658     Biso  1.000000 C
   C15        1.0     0.923135      0.790783      0.600508     Biso  1.000000 C
   C16        1.0     0.819650      0.727399      0.469794     Biso  1.000000 C
   C17        1.0     0.068571      0.783612      0.584141     Biso  1.000000 C
   C18        1.0     0.868938      0.654927      0.319429     Biso  1.000000 C
   C19        1.0     0.121857      0.708619      0.432794     Biso  1.000000 C
   C20        1.0     0.013519      0.644539      0.296276     Biso  1.000000 C
   C21        1.0     0.266191      0.696184      0.413892     Biso  1.000000 C
   C22        1.0     0.155296      0.407668      0.839435     Biso  1.000000 C
   C23        1.0     0.010326      0.425958      0.849979     Biso  1.000000 C
   C24        1.0     0.254255      0.467876      0.971836     Biso  1.000000 C
   C25        1.0     0.967179      0.505214      0.001333     Biso  1.000000 C
   C26        1.0     0.214710      0.554139      0.127199     Biso  1.000000 C
   C27        1.0     0.061690      0.566061      0.138509     Biso  1.000000 C
   C28        1.0     0.313113      0.621339      0.260595     Biso  1.000000 C
   H1         1.0     0.584842      0.403026      0.751999     Biso  1.000000 H
   H2         1.0     0.582287      0.703831      0.723173     Biso  1.000000 H
   H3         1.0     0.636165      0.148321      0.709551     Biso  1.000000 H
   H4         1.0     0.554434      0.474137      0.528113     Biso  1.000000 H
   H5         1.0     0.595389      0.423431      0.244908     Biso  1.000000 H
   H6         1.0     0.629981      0.959776      0.436328     Biso  1.000000 H
   H7         1.0     0.596769      0.621546      0.946389     Biso  1.000000 H
   H8         1.0     0.566603      0.964921      0.773787     Biso  1.000000 H
   H9         1.0     0.583554      0.233814      0.965319     Biso  1.000000 H
   H10        1.0     0.587515      0.806229      0.218040     Biso  1.000000 H
   H11        1.0     0.888685      0.841818      0.718263     Biso  1.000000 H
   H12        1.0     0.191271      0.346975      0.724560     Biso  1.000000 H
   H13        1.0     0.705388      0.728894      0.482788     Biso  1.000000 H
   H14        1.0     0.146079      0.832761      0.688773     Biso  1.000000 H
   H15        1.0     0.343916      0.738446      0.518119     Biso  1.000000 H
   H16        1.0     0.788390      0.602410      0.218057     Biso  1.000000 H
   H17        1.0     0.930571      0.374521      0.747868     Biso  1.000000 H
   H18        1.0     0.366823      0.454261      0.959659     Biso  1.000000 H
   H19        1.0     0.426654      0.611949      0.248734     Biso  1.000000 H
   H20        1.0     0.854352      0.518076      0.008976     Biso  1.000000 H
   K1         1.0     0.288439      0.302419      0.229998     Biso  1.000000 K
   K2         1.0     0.895456      0.943764      0.273912     Biso  1.000000 K
   K3         1.0     0.888073      0.164401      0.939299     Biso  1.000000 K
   K4         1.0     0.275117      0.779407      0.978011     Biso  1.000000 K
   K5         1.0     0.898250      0.445090      0.481604     Biso  1.000000 K
   K6         1.0     0.305167      0.035532      0.514921     Biso  1.000000 K
\end{verbatim}
\end{tiny}

\subsubsection{Structure \textbf{EA3(M)}}
\begin{tiny}
\begin{verbatim}
#======================================================================

# CRYSTAL DATA

#----------------------------------------------------------------------

STRUCTURE_EA3M


_pd_phase_name                         'EA363  10.231  9.177  9.795 96.002 106'
_cell_length_a                         10.20131
_cell_length_b                         9.12298
_cell_length_c                         9.59360
_cell_angle_alpha                      97.24651
_cell_angle_beta                       107.84360
_cell_angle_gamma                      116.23448
_symmetry_space_group_name_H-M         'P 1'
_symmetry_Int_Tables_number            1

loop_
_symmetry_equiv_pos_as_xyz
   'x, y, z'

loop_
   _atom_site_label
   _atom_site_occupancy
   _atom_site_fract_x
   _atom_site_fract_y
   _atom_site_fract_z
   _atom_site_adp_type
   _atom_site_B_iso_or_equiv
   _atom_site_type_symbol
   C1         1.0     0.836169      0.109741      0.105015     Biso  1.000000 C
   C2         1.0     0.834416      0.243207      0.046033     Biso  1.000000 C
   C3         1.0     0.746000      0.044297      0.191423     Biso  1.000000 C
   C4         1.0     0.735750      0.307443      0.071244     Biso  1.000000 C
   C5         1.0     0.646497      0.106824      0.220892     Biso  1.000000 C
   C6         1.0     0.633995      0.236727      0.147982     Biso  1.000000 C
   C7         1.0     0.561124      0.052875      0.312388     Biso  1.000000 C
   C8         1.0     0.297715      0.394569      0.174598     Biso  1.000000 C
   C9         1.0     0.372780      0.442383      0.073868     Biso  1.000000 C
   C10        1.0     0.318861      0.281037      0.254838     Biso  1.000000 C
   C11        1.0     0.470581      0.375044      0.054245     Biso  1.000000 C
   C12        1.0     0.427385      0.222690      0.248027     Biso  1.000000 C
   C13        1.0     0.507925      0.274398      0.143887     Biso  1.000000 C
   C14        1.0     0.463002      0.122565      0.336438     Biso  1.000000 C
   C15        1.0     0.344525      0.621126      0.613722     Biso  1.000000 C
   C16        1.0     0.329814      0.740943      0.540592     Biso  1.000000 C
   C17        1.0     0.256285      0.558847      0.702118     Biso  1.000000 C
   C18        1.0     0.222949      0.797038      0.556454     Biso  1.000000 C
   C19        1.0     0.149079      0.614115      0.723134     Biso  1.000000 C
   C20        1.0     0.129437      0.735750      0.642254     Biso  1.000000 C
   C21        1.0     0.065917      0.561480      0.816619     Biso  1.000000 C
   C22        1.0     0.780823      0.879139      0.663294     Biso  1.000000 C
   C23        1.0     0.873317      0.949807      0.581274     Biso  1.000000 C
   C24        1.0     0.800106      0.760620      0.737394     Biso  1.000000 C
   C25        1.0     0.981715      0.895235      0.568793     Biso  1.000000 C
   C26        1.0     0.922675      0.720091      0.743486     Biso  1.000000 C
   C27        1.0     0.010051      0.782195      0.646685     Biso  1.000000 C
   C28        1.0     0.963751      0.627004      0.837422     Biso  1.000000 C
   H1         1.0     0.909430      0.057127      0.087055     Biso  1.000000 H
   H2         1.0     0.217092      0.438509      0.188195     Biso  1.000000 H
   H3         1.0     0.911823      0.300989      0.987497     Biso  1.000000 H
   H4         1.0     0.750406      0.941652      0.236699     Biso  1.000000 H
   H5         1.0     0.571257      0.957965      0.367768     Biso  1.000000 H
   H6         1.0     0.735216      0.411721      0.026397     Biso  1.000000 H
   H7         1.0     0.352770      0.524115      0.006532     Biso  1.000000 H
   H8         1.0     0.260751      0.246370      0.333778     Biso  1.000000 H
   H9         1.0     0.400061      0.081364      0.411215     Biso  1.000000 H
   H10        1.0     0.526383      0.411418      0.973989     Biso  1.000000 H
   H11        1.0     0.426017      0.577094      0.604059     Biso  1.000000 H
   H12        1.0     0.689907      0.910349      0.667575     Biso  1.000000 H
   H13        1.0     0.403878      0.794981      0.478233     Biso  1.000000 H
   H14        1.0     0.269122      0.465385      0.757293     Biso  1.000000 H
   H15        1.0     0.082444      0.472639      0.877286     Biso  1.000000 H
   H16        1.0     0.213492      0.891304      0.500615     Biso  1.000000 H
   H17        1.0     0.856440      0.036942      0.519337     Biso  1.000000 H
   H18        1.0     0.730930      0.710835      0.804949     Biso  1.000000 H
   H19        1.0     0.896508      0.580708      0.907857     Biso  1.000000 H
   H20        1.0     0.049128      0.948377      0.500941     Biso  1.000000 H
   K1         1.0     0.219902      0.766625      0.154599     Biso  1.000000 K
   K2         1.0     0.651793      0.599920      0.378940     Biso  1.000000 K
   K3         1.0     0.718098      0.282018      0.649843     Biso  1.000000 K
   K4         1.0     0.486870      0.931263      0.880648     Biso  1.000000 K
   K5         1.0     0.122075      0.088187      0.904903     Biso  1.000000 K
   K6         1.0     0.996509      0.406150      0.389058     Biso  1.000000 K
\end{verbatim}
\end{tiny}

\subsubsection{Structure \textbf{EA3(SM)}}
\begin{tiny}
\begin{verbatim}
#======================================================================

# CRYSTAL DATA

#----------------------------------------------------------------------

STRUCTURE_EA3SM


_pd_phase_name                         'EA1363  9.952  8.912  9.476 95.049 114'
_cell_length_a                         9.96513
_cell_length_b                         8.92428
_cell_length_c                         9.48859
_cell_angle_alpha                      95.07159
_cell_angle_beta                       114.24680
_cell_angle_gamma                      64.96310
_symmetry_space_group_name_H-M         'P 1'
_symmetry_Int_Tables_number            1

loop_
_symmetry_equiv_pos_as_xyz
   'x, y, z'

loop_
   _atom_site_label
   _atom_site_occupancy
   _atom_site_fract_x
   _atom_site_fract_y
   _atom_site_fract_z
   _atom_site_adp_type
   _atom_site_B_iso_or_equiv
   _atom_site_type_symbol
   C1         1.0     0.188870      0.490867      0.617449     Biso  1.000000 C
   C2         1.0     0.200130      0.630211      0.572191     Biso  1.000000 C
   C3         1.0     0.099609      0.507258      0.705628     Biso  1.000000 C
   C4         1.0     0.113623      0.790365      0.612481     Biso  1.000000 C
   C5         1.0     0.009944      0.666282      0.747380     Biso  1.000000 C
   C6         1.0     0.008819      0.814873      0.687234     Biso  1.000000 C
   C7         1.0     0.922491      0.689236      0.837597     Biso  1.000000 C
   C8         1.0     0.707270      0.306014      0.756389     Biso  1.000000 C
   C9         1.0     0.773794      0.289074      0.647278     Biso  1.000000 C
   C10        1.0     0.720710      0.164395      0.827448     Biso  1.000000 C
   C11        1.0     0.859100      0.126408      0.613167     Biso  1.000000 C
   C12        1.0     0.812035      0.998054      0.801361     Biso  1.000000 C
   C13        1.0     0.889388      0.979432      0.693220     Biso  1.000000 C
   C14        1.0     0.834865      0.853763      0.875005     Biso  1.000000 C
   C15        1.0     0.722862      0.492121      0.151052     Biso  1.000000 C
   C16        1.0     0.694191      0.631062      0.064311     Biso  1.000000 C
   C17        1.0     0.635820      0.510467      0.241913     Biso  1.000000 C
   C18        1.0     0.577082      0.791924      0.071942     Biso  1.000000 C
   C19        1.0     0.522097      0.670823      0.258024     Biso  1.000000 C
   C20        1.0     0.496798      0.818160      0.172331     Biso  1.000000 C
   C21        1.0     0.438289      0.694419      0.352120     Biso  1.000000 C
   C22        1.0     0.166688      0.309117      0.214225     Biso  1.000000 C
   C23        1.0     0.269345      0.294044      0.141387     Biso  1.000000 C
   C24        1.0     0.184732      0.167341      0.288923     Biso  1.000000 C
   C25        1.0     0.386809      0.132244      0.138699     Biso  1.000000 C
   C26        1.0     0.301140      0.002229      0.288186     Biso  1.000000 C
   C27        1.0     0.400164      0.984596      0.201381     Biso  1.000000 C
   C28        1.0     0.323202      0.859139      0.362391     Biso  1.000000 C
   H1         1.0     0.248359      0.367999      0.584192     Biso  1.000000 H
   H2         1.0     0.646316      0.428107      0.790190     Biso  1.000000 H
   H3         1.0     0.281099      0.614903      0.516386     Biso  1.000000 H
   H4         1.0     0.091083      0.397785      0.737424     Biso  1.000000 H
   H5         1.0     0.925866      0.578693      0.884020     Biso  1.000000 H
   H6         1.0     0.119290      0.899106      0.574889     Biso  1.000000 H
   H7         1.0     0.761475      0.397249      0.588944     Biso  1.000000 H
   H8         1.0     0.670542      0.179496      0.913754     Biso  1.000000 H
   H9         1.0     0.775669      0.868142      0.953792     Biso  1.000000 H
   H10        1.0     0.911561      0.114718      0.529504     Biso  1.000000 H
   H11        1.0     0.812182      0.368802      0.147367     Biso  1.000000 H
   H12        1.0     0.067110      0.428003      0.210372     Biso  1.000000 H
   H13        1.0     0.764121      0.615999      0.996470     Biso  1.000000 H
   H14        1.0     0.659632      0.400959      0.308117     Biso  1.000000 H
   H15        1.0     0.459563      0.584298      0.416322     Biso  1.000000 H
   H16        1.0     0.558615      0.900147      0.008019     Biso  1.000000 H
   H17        1.0     0.252569      0.400247      0.076164     Biso  1.000000 H
   H18        1.0     0.104978      0.181809      0.346102     Biso  1.000000 H
   H19        1.0     0.248747      0.873786      0.426008     Biso  1.000000 H
   H20        1.0     0.463124      0.122371      0.079128     Biso  1.000000 H
   K1         1.0     0.569192      0.755658      0.687538     Biso  1.000000 K
   K2         1.0     0.375475      0.587892      0.921694     Biso  1.000000 K
   K3         1.0     0.506275      0.189168      0.476439     Biso  1.000000 K
   K4         1.0     0.107961      0.703157      0.180947     Biso  1.000000 K
   K5         1.0     0.070403      0.125132      0.941596     Biso  1.000000 K
   K6         1.0     0.838530      0.696998      0.402140     Biso  1.000000 K
\end{verbatim}
\end{tiny}

\subsubsection{Structure \textbf{PS--K$_3$Phenanthrene}}
\begin{tiny}
\begin{verbatim}
#======================================================================

# CRYSTAL DATA

#----------------------------------------------------------------------

STRUCTURE_PS_K3PA


_pd_phase_name                         'Structure from PWSCF output 17438TITLE'
_cell_length_a                         9.17265
_cell_length_b                         6.51744
_cell_length_c                         10.35332
_cell_angle_alpha                      90
_cell_angle_beta                       102.60990
_cell_angle_gamma                      90
_symmetry_space_group_name_H-M         'P 1'
_symmetry_Int_Tables_number            1

loop_
_symmetry_equiv_pos_as_xyz
   'x, y, z'

loop_
   _atom_site_label
   _atom_site_occupancy
   _atom_site_fract_x
   _atom_site_fract_y
   _atom_site_fract_z
   _atom_site_adp_type
   _atom_site_B_iso_or_equiv
   _atom_site_type_symbol
   C1         1.0     0.750724      0.105397      0.250145     Biso  1.000000 C
   C2         1.0     0.249276      0.605397      0.749855     Biso  1.000000 C
   C3         1.0     0.884977      0.013301      0.323302     Biso  1.000000 C
   C4         1.0     0.115023      0.513301      0.676698     Biso  1.000000 C
   C5         1.0     0.973827      0.897250      0.257254     Biso  1.000000 C
   C6         1.0     0.026173      0.397250      0.742746     Biso  1.000000 C
   C7         1.0     0.928428      0.870674      0.117023     Biso  1.000000 C
   C8         1.0     0.071572      0.370674      0.882976     Biso  1.000000 C
   C9         1.0     0.848052      0.851466      0.818369     Biso  1.000000 C
   C10        1.0     0.151948      0.351466      0.181631     Biso  1.000000 C
   C11        1.0     0.817825      0.860547      0.677282     Biso  1.000000 C
   C12        1.0     0.182175      0.360547      0.322718     Biso  1.000000 C
   C13        1.0     0.688699      0.959540      0.607544     Biso  1.000000 C
   C14        1.0     0.311301      0.459540      0.392456     Biso  1.000000 C
   C15        1.0     0.588481      0.049775      0.678163     Biso  1.000000 C
   C16        1.0     0.411519      0.549775      0.321837     Biso  1.000000 C
   C17        1.0     0.535092      0.170798      0.893078     Biso  1.000000 C
   C18        1.0     0.464909      0.670798      0.106922     Biso  1.000000 C
   C19        1.0     0.571244      0.175743      0.035624     Biso  1.000000 C
   C20        1.0     0.428756      0.675743      0.964376     Biso  1.000000 C
   C21        1.0     0.702215      0.084919      0.108800     Biso  1.000000 C
   C22        1.0     0.297785      0.584919      0.891200     Biso  1.000000 C
   C23        1.0     0.798121      0.967964      0.038260     Biso  1.000000 C
   C24        1.0     0.201879      0.467964      0.961740     Biso  1.000000 C
   C25        1.0     0.758046      0.955764      0.894669     Biso  1.000000 C
   C26        1.0     0.241954      0.455764      0.105330     Biso  1.000000 C
   C27        1.0     0.622781      0.062147      0.820924     Biso  1.000000 C
   C28        1.0     0.377219      0.562147      0.179076     Biso  1.000000 C
   H1         1.0     0.681901      0.197403      0.302743     Biso  1.000000 H
   H2         1.0     0.318099      0.697403      0.697257     Biso  1.000000 H
   H3         1.0     0.913669      0.032517      0.431032     Biso  1.000000 H
   H4         1.0     0.086331      0.532517      0.568968     Biso  1.000000 H
   H5         1.0     0.076771      0.823467      0.310402     Biso  1.000000 H
   H6         1.0     0.923229      0.323467      0.689598     Biso  1.000000 H
   H7         1.0     0.000830      0.777365      0.069427     Biso  1.000000 H
   H8         1.0     0.999170      0.277365      0.930573     Biso  1.000000 H
   H9         1.0     0.949378      0.770712      0.867420     Biso  1.000000 H
   H10        1.0     0.050622      0.270713      0.132580     Biso  1.000000 H
   H11        1.0     0.894871      0.784578      0.625862     Biso  1.000000 H
   H12        1.0     0.105129      0.284578      0.374138     Biso  1.000000 H
   H13        1.0     0.668494      0.967945      0.499535     Biso  1.000000 H
   H14        1.0     0.331506      0.467945      0.500465     Biso  1.000000 H
   H15        1.0     0.488337      0.129799      0.623469     Biso  1.000000 H
   H16        1.0     0.511663      0.629799      0.376531     Biso  1.000000 H
   H17        1.0     0.435468      0.253438      0.840658     Biso  1.000000 H
   H18        1.0     0.564532      0.753438      0.159342     Biso  1.000000 H
   H19        1.0     0.508199      0.280162      0.085588     Biso  1.000000 H
   H20        1.0     0.491801      0.780162      0.914412     Biso  1.000000 H
   K1         1.0     0.833687      0.553203      0.354523     Biso  1.000000 K
   K2         1.0     0.166313      0.053203      0.645477     Biso  1.000000 K
   K3         1.0     0.770137      0.495779      0.002727     Biso  1.000000 K
   K4         1.0     0.229863      0.995779      0.997273     Biso  1.000000 K
   K5         1.0     0.582063      0.548047      0.691408     Biso  1.000000 K
   K6         1.0     0.417937      0.048047      0.308593     Biso  1.000000 K
\end{verbatim}
\end{tiny}

\vspace{15cm}
\subsection{K$_3$-picene}

\subsubsection{Structure \textbf{A}}
\begin{tiny}
\begin{verbatim}
#======================================================================

# CRYSTAL DATA

#----------------------------------------------------------------------

STRUCTURE_A


_pd_phase_name                         'EA277  10.292  7.529 12.847 76.857 90.'
_cell_length_a                         10.01017
_cell_length_b                         7.32203
_cell_length_c                         12.67017
_cell_angle_alpha                      77.62717
_cell_angle_beta                       90.21521
_cell_angle_gamma                      90.24959
_symmetry_space_group_name_H-M         'P 1'
_symmetry_Int_Tables_number            1

loop_
_symmetry_equiv_pos_as_xyz
   'x, y, z'

loop_
   _atom_site_label
   _atom_site_occupancy
   _atom_site_fract_x
   _atom_site_fract_y
   _atom_site_fract_z
   _atom_site_adp_type
   _atom_site_B_iso_or_equiv
   _atom_site_type_symbol
   C1         1.0     0.043654      0.976564      0.361841     Biso  1.000000 C
   C2         1.0     0.081781      0.877539      0.282139     Biso  1.000000 C
   C3         1.0     0.078243      0.912305      0.473544     Biso  1.000000 C
   C4         1.0     0.141312      0.692073      0.316783     Biso  1.000000 C
   C5         1.0     0.069358      0.954650      0.166345     Biso  1.000000 C
   C6         1.0     0.174209      0.629734      0.429822     Biso  1.000000 C
   C7         1.0     0.163292      0.585507      0.237452     Biso  1.000000 C
   C8         1.0     0.146225      0.742327      0.505583     Biso  1.000000 C
   C9         1.0     0.067339      0.828101      0.090218     Biso  1.000000 C
   C10        1.0     0.118259      0.647956      0.128066     Biso  1.000000 C
   C11        1.0     0.017623      0.900508      0.979890     Biso  1.000000 C
   C12        1.0     0.051659      0.147592      0.124559     Biso  1.000000 C
   C13        1.0     0.008390      0.217869      0.018269     Biso  1.000000 C
   C14        1.0     0.981342      0.096498      0.946359     Biso  1.000000 C
   C15        1.0     0.921403      0.164994      0.841494     Biso  1.000000 C
   C16        1.0     0.005607      0.787369      0.903359     Biso  1.000000 C
   C17        1.0     0.908775      0.040478      0.766241     Biso  1.000000 C
   C18        1.0     0.955701      0.854826      0.798730     Biso  1.000000 C
   C19        1.0     0.848008      0.109476      0.662896     Biso  1.000000 C
   C20        1.0     0.871068      0.352253      0.807458     Biso  1.000000 C
   C21        1.0     0.811134      0.415877      0.704647     Biso  1.000000 C
   C22        1.0     0.799886      0.294420      0.632570     Biso  1.000000 C
   C23        1.0     0.543177      0.512665      0.140261     Biso  1.000000 C
   C24        1.0     0.581939      0.611490      0.220209     Biso  1.000000 C
   C25        1.0     0.577626      0.577539      0.028688     Biso  1.000000 C
   C26        1.0     0.641963      0.797226      0.186134     Biso  1.000000 C
   C27        1.0     0.569073      0.534004      0.335984     Biso  1.000000 C
   C28        1.0     0.675657      0.859541      0.073334     Biso  1.000000 C
   C29        1.0     0.663216      0.904213      0.265506     Biso  1.000000 C
   C30        1.0     0.646763      0.747390      0.997320     Biso  1.000000 C
   C31        1.0     0.567177      0.660418      0.412108     Biso  1.000000 C
   C32        1.0     0.618061      0.841165      0.374652     Biso  1.000000 C
   C33        1.0     0.517561      0.587139      0.522280     Biso  1.000000 C
   C34        1.0     0.551172      0.341009      0.377371     Biso  1.000000 C
   C35        1.0     0.507594      0.269999      0.483380     Biso  1.000000 C
   C36        1.0     0.480885      0.390907      0.555290     Biso  1.000000 C
   C37        1.0     0.420937      0.321573      0.659868     Biso  1.000000 C
   C38        1.0     0.506096      0.699626      0.599043     Biso  1.000000 C
   C39        1.0     0.409193      0.445347      0.735498     Biso  1.000000 C
   C40        1.0     0.456505      0.631289      0.703492     Biso  1.000000 C
   C41        1.0     0.348640      0.375104      0.838624     Biso  1.000000 C
   C42        1.0     0.369861      0.134179      0.693136     Biso  1.000000 C
   C43        1.0     0.310013      0.069443      0.795611     Biso  1.000000 C
   C44        1.0     0.299652      0.190155      0.868119     Biso  1.000000 C
   H1         1.0     0.990681      0.109126      0.337002     Biso  1.000000 H
   H2         1.0     0.053060      0.995470      0.532408     Biso  1.000000 H
   H3         1.0     0.234078      0.502795      0.453402     Biso  1.000000 H
   H4         1.0     0.209082      0.447498      0.262070     Biso  1.000000 H
   H5         1.0     0.183762      0.702932      0.588538     Biso  1.000000 H
   H6         1.0     0.131076      0.552972      0.073021     Biso  1.000000 H
   H7         1.0     0.073601      0.244154      0.177459     Biso  1.000000 H
   H8         1.0     0.008935      0.368109      0.985571     Biso  1.000000 H
   H9         1.0     0.039392      0.642720      0.923935     Biso  1.000000 H
   H10        1.0     0.948486      0.761894      0.742098     Biso  1.000000 H
   H11        1.0     0.836106      0.014331      0.607825     Biso  1.000000 H
   H12        1.0     0.873084      0.446962      0.863002     Biso  1.000000 H
   H13        1.0     0.767248      0.555438      0.684253     Biso  1.000000 H
   H14        1.0     0.744649      0.335233      0.556945     Biso  1.000000 H
   H15        1.0     0.489432      0.380186      0.164712     Biso  1.000000 H
   H16        1.0     0.551735      0.494776      0.969516     Biso  1.000000 H
   H17        1.0     0.736135      0.986560      0.050115     Biso  1.000000 H
   H18        1.0     0.709049      0.042698      0.241228     Biso  1.000000 H
   H19        1.0     0.684643      0.786849      0.914642     Biso  1.000000 H
   H20        1.0     0.630755      0.936242      0.429788     Biso  1.000000 H
   H21        1.0     0.573030      0.244887      0.324400     Biso  1.000000 H
   H22        1.0     0.507933      0.119773      0.516023     Biso  1.000000 H
   H23        1.0     0.540157      0.844580      0.578801     Biso  1.000000 H
   H24        1.0     0.449694      0.723902      0.760291     Biso  1.000000 H
   H25        1.0     0.337260      0.469618      0.894015     Biso  1.000000 H
   H26        1.0     0.371010      0.040292      0.637102     Biso  1.000000 H
   H27        1.0     0.265521      0.929804      0.815439     Biso  1.000000 H
   H28        1.0     0.244365      0.148477      0.943427     Biso  1.000000 H
   K1         1.0     0.366052      0.853274      0.080021     Biso  1.000000 K
   K2         1.0     0.865147      0.636019      0.422516     Biso  1.000000 K
   K3         1.0     0.613925      0.141035      0.831662     Biso  1.000000 K
   K4         1.0     0.113814      0.345484      0.669154     Biso  1.000000 K
   K5         1.0     0.847151      0.495566      0.129264     Biso  1.000000 K
   K6         1.0     0.347314      0.991941      0.372045     Biso  1.000000 K
\end{verbatim}
\end{tiny}

\subsubsection{Structure \textbf{B}}
\begin{tiny}
\begin{verbatim}
#======================================================================

# CRYSTAL DATA

#----------------------------------------------------------------------

STRUCTURE_B


_pd_phase_name                         'EA334  10.672  7.649 14.575 91.235 105'
_cell_length_a                         9.83379
_cell_length_b                         6.91916
_cell_length_c                         13.86159
_cell_angle_alpha                      86.02428
_cell_angle_beta                       100.67679
_cell_angle_gamma                      78.55915
_symmetry_space_group_name_H-M         'P 1'
_symmetry_Int_Tables_number            1

loop_
_symmetry_equiv_pos_as_xyz
   'x, y, z'

loop_
   _atom_site_label
   _atom_site_occupancy
   _atom_site_fract_x
   _atom_site_fract_y
   _atom_site_fract_z
   _atom_site_adp_type
   _atom_site_B_iso_or_equiv
   _atom_site_type_symbol
   C1         1.0     0.552595      0.880908      0.168771     Biso  1.000000 C
   C2         1.0     0.473584      0.737718      0.131609     Biso  1.000000 C
   C3         1.0     0.600430      0.885843      0.271189     Biso  1.000000 C
   C4         1.0     0.457112      0.585304      0.204640     Biso  1.000000 C
   C5         1.0     0.414016      0.734769      0.028593     Biso  1.000000 C
   C6         1.0     0.500963      0.599711      0.307125     Biso  1.000000 C
   C7         1.0     0.403653      0.421083      0.168787     Biso  1.000000 C
   C8         1.0     0.573260      0.749724      0.341199     Biso  1.000000 C
   C9         1.0     0.349410      0.567716      0.995093     Biso  1.000000 C
   C10        1.0     0.351769      0.414275      0.068532     Biso  1.000000 C
   C11        1.0     0.296296      0.560712      0.890712     Biso  1.000000 C
   C12        1.0     0.423701      0.881770      0.953227     Biso  1.000000 C
   C13        1.0     0.377486      0.870766      0.852644     Biso  1.000000 C
   C14        1.0     0.312782      0.713113      0.816287     Biso  1.000000 C
   C15        1.0     0.270559      0.696862      0.712782     Biso  1.000000 C
   C16        1.0     0.233012      0.406125      0.854465     Biso  1.000000 C
   C17        1.0     0.192755      0.544446      0.680282     Biso  1.000000 C
   C18        1.0     0.180969      0.398783      0.754451     Biso  1.000000 C
   C19        1.0     0.142772      0.537078      0.577919     Biso  1.000000 C
   C20        1.0     0.306503      0.817543      0.635841     Biso  1.000000 C
   C21        1.0     0.263572      0.796468      0.534952     Biso  1.000000 C
   C22        1.0     0.179120      0.660866      0.504242     Biso  1.000000 C
   C23        1.0     0.695282      0.986673      0.682902     Biso  1.000000 C
   C24        1.0     0.753680      0.148604      0.720077     Biso  1.000000 C
   C25        1.0     0.644108      0.982125      0.580914     Biso  1.000000 C
   C26        1.0     0.763072      0.304956      0.646581     Biso  1.000000 C
   C27        1.0     0.796466      0.167512      0.823638     Biso  1.000000 C
   C28        1.0     0.716558      0.291041      0.544214     Biso  1.000000 C
   C29        1.0     0.813168      0.472219      0.681519     Biso  1.000000 C
   C30        1.0     0.653726      0.132028      0.510649     Biso  1.000000 C
   C31        1.0     0.867384      0.329376      0.855858     Biso  1.000000 C
   C32        1.0     0.868026      0.479001      0.781286     Biso  1.000000 C
   C33        1.0     0.919241      0.339281      0.959941     Biso  1.000000 C
   C34        1.0     0.770680      0.037106      0.900116     Biso  1.000000 C
   C35        1.0     0.817797      0.049011      0.000275     Biso  1.000000 C
   C36        1.0     0.899150      0.191909      0.035365     Biso  1.000000 C
   C37        1.0     0.957127      0.196181      0.137999     Biso  1.000000 C
   C38        1.0     0.981297      0.495859      0.995345     Biso  1.000000 C
   C39        1.0     0.028260      0.356152      0.169400     Biso  1.000000 C
   C40        1.0     0.032023      0.506669      0.094861     Biso  1.000000 C
   C41        1.0     0.082823      0.363170      0.271056     Biso  1.000000 C
   C42        1.0     0.943665      0.055374      0.216192     Biso  1.000000 C
   C43        1.0     0.992869      0.073314      0.316678     Biso  1.000000 C
   C44        1.0     0.063461      0.224852      0.346005     Biso  1.000000 C
   H1         1.0     0.574969      0.991451      0.117594     Biso  1.000000 H
   H2         1.0     0.661264      0.996092      0.295348     Biso  1.000000 H
   H3         1.0     0.484638      0.486072      0.359946     Biso  1.000000 H
   H4         1.0     0.402734      0.299343      0.222996     Biso  1.000000 H
   H5         1.0     0.611890      0.753147      0.419807     Biso  1.000000 H
   H6         1.0     0.312936      0.283385      0.046685     Biso  1.000000 H
   H7         1.0     0.469269      0.007950      0.974472     Biso  1.000000 H
   H8         1.0     0.386046      0.991418      0.800742     Biso  1.000000 H
   H9         1.0     0.221362      0.289104      0.906883     Biso  1.000000 H
   H10        1.0     0.129318      0.278783      0.731088     Biso  1.000000 H
   H11        1.0     0.081516      0.425748      0.555832     Biso  1.000000 H
   H12        1.0     0.377427      0.917373      0.656328     Biso  1.000000 H
   H13        1.0     0.298304      0.886988      0.480427     Biso  1.000000 H
   H14        1.0     0.143053      0.648800      0.426132     Biso  1.000000 H
   H15        1.0     0.686472      0.865319      0.734051     Biso  1.000000 H
   H16        1.0     0.595271      0.859835      0.557754     Biso  1.000000 H
   H17        1.0     0.722616      0.412321      0.491457     Biso  1.000000 H
   H18        1.0     0.815905      0.592410      0.627330     Biso  1.000000 H
   H19        1.0     0.612400      0.128879      0.432261     Biso  1.000000 H
   H20        1.0     0.910507      0.607859      0.801434     Biso  1.000000 H
   H21        1.0     0.707884      0.927106      0.879647     Biso  1.000000 H
   H22        1.0     0.791554      0.945734      0.053207     Biso  1.000000 H
   H23        1.0     0.990579      0.612268      0.941539     Biso  1.000000 H
   H24        1.0     0.079687      0.630307      0.118022     Biso  1.000000 H
   H25        1.0     0.136649      0.482800      0.291569     Biso  1.000000 H
   H26        1.0     0.889524      0.935718      0.197671     Biso  1.000000 H
   H27        1.0     0.977369      0.964024      0.371980     Biso  1.000000 H
   H28        1.0     0.105677      0.233974      0.423716     Biso  1.000000 H
   K1         1.0     0.273239      0.975336      0.245553     Biso  1.000000 K
   K2         1.0     0.768464      0.470129      0.243626     Biso  1.000000 K
   K3         1.0     0.611149      0.480424      0.921872     Biso  1.000000 K
   K4         1.0     0.100064      0.940449      0.928312     Biso  1.000000 K
   K5         1.0     0.970566      0.946548      0.592755     Biso  1.000000 K
   K6         1.0     0.449855      0.360329      0.621226     Biso  1.000000 K
\end{verbatim}
\end{tiny}

\subsubsection{Structure \textbf{C}}
\begin{tiny}
\begin{verbatim}
#======================================================================

# CRYSTAL DATA

#----------------------------------------------------------------------

STRUCTURE_C


_pd_phase_name                         'EA188   7.932  7.540 13.623 97.445 89.'
_cell_length_a                         7.93912
_cell_length_b                         7.55680
_cell_length_c                         13.65582
_cell_angle_alpha                      97.43658
_cell_angle_beta                       89.93323
_cell_angle_gamma                      90.02206
_symmetry_space_group_name_H-M         'P 1'
_symmetry_Int_Tables_number            1

loop_
_symmetry_equiv_pos_as_xyz
   'x, y, z'

loop_
   _atom_site_label
   _atom_site_occupancy
   _atom_site_fract_x
   _atom_site_fract_y
   _atom_site_fract_z
   _atom_site_adp_type
   _atom_site_B_iso_or_equiv
   _atom_site_type_symbol
   C1         1.0     0.576489      0.657849      0.595051     Biso  1.000000 C
   C2         1.0     0.459376      0.715317      0.528319     Biso  1.000000 C
   C3         1.0     0.684022      0.514905      0.564808     Biso  1.000000 C
   C4         1.0     0.450372      0.633778      0.427577     Biso  1.000000 C
   C5         1.0     0.676938      0.430540      0.465591     Biso  1.000000 C
   C6         1.0     0.563897      0.485830      0.394806     Biso  1.000000 C
   C7         1.0     0.340178      0.696032      0.358088     Biso  1.000000 C
   C8         1.0     0.561066      0.404926      0.290909     Biso  1.000000 C
   C9         1.0     0.337059      0.614361      0.258149     Biso  1.000000 C
   C10        1.0     0.442908      0.473180      0.221852     Biso  1.000000 C
   C11        1.0     0.673859      0.268407      0.252602     Biso  1.000000 C
   C12        1.0     0.443740      0.396212      0.115872     Biso  1.000000 C
   C13        1.0     0.679198      0.200211      0.150556     Biso  1.000000 C
   C14        1.0     0.564513      0.255002      0.081313     Biso  1.000000 C
   C15        1.0     0.336591      0.458499      0.046588     Biso  1.000000 C
   C16        1.0     0.567367      0.180169      0.976436     Biso  1.000000 C
   C17        1.0     0.343995      0.392719      0.944076     Biso  1.000000 C
   C18        1.0     0.675806      0.038310      0.937540     Biso  1.000000 C
   C19        1.0     0.460023      0.258767      0.906976     Biso  1.000000 C
   C20        1.0     0.698893      0.990447      0.833583     Biso  1.000000 C
   C21        1.0     0.481638      0.203158      0.802941     Biso  1.000000 C
   C22        1.0     0.602461      0.074934      0.767352     Biso  1.000000 C
   C23        1.0     0.081111      0.595351      0.742062     Biso  1.000000 C
   C24        1.0     0.963088      0.537403      0.808277     Biso  1.000000 C
   C25        1.0     0.188704      0.737923      0.772899     Biso  1.000000 C
   C26        1.0     0.953210      0.617848      0.909228     Biso  1.000000 C
   C27        1.0     0.180622      0.821248      0.872301     Biso  1.000000 C
   C28        1.0     0.066502      0.765640      0.942487     Biso  1.000000 C
   C29        1.0     0.842280      0.554983      0.978279     Biso  1.000000 C
   C30        1.0     0.062183      0.846589      0.046383     Biso  1.000000 C
   C31        1.0     0.838539      0.635790      0.078406     Biso  1.000000 C
   C32        1.0     0.943876      0.777179      0.115060     Biso  1.000000 C
   C33        1.0     0.173905      0.983825      0.084856     Biso  1.000000 C
   C34        1.0     0.943795      0.853642      0.221104     Biso  1.000000 C
   C35        1.0     0.178848      0.050998      0.187021     Biso  1.000000 C
   C36        1.0     0.063983      0.994998      0.255970     Biso  1.000000 C
   C37        1.0     0.836356      0.790335      0.290110     Biso  1.000000 C
   C38        1.0     0.066458      0.068779      0.360992     Biso  1.000000 C
   C39        1.0     0.843019      0.855477      0.392678     Biso  1.000000 C
   C40        1.0     0.175517      0.209734      0.400293     Biso  1.000000 C
   C41        1.0     0.958610      0.989578      0.430063     Biso  1.000000 C
   C42        1.0     0.198193      0.256449      0.504330     Biso  1.000000 C
   C43        1.0     0.979375      0.044550      0.534169     Biso  1.000000 C
   C44        1.0     0.100486      0.172020      0.570156     Biso  1.000000 C
   H1         1.0     0.587479      0.733203      0.668356     Biso  1.000000 H
   H2         1.0     0.375724      0.828039      0.553226     Biso  1.000000 H
   H3         1.0     0.772037      0.466263      0.616706     Biso  1.000000 H
   H4         1.0     0.767949      0.324133      0.443663     Biso  1.000000 H
   H5         1.0     0.257104      0.809641      0.379862     Biso  1.000000 H
   H6         1.0     0.250908      0.673092      0.208955     Biso  1.000000 H
   H7         1.0     0.768532      0.218367      0.300258     Biso  1.000000 H
   H8         1.0     0.776693      0.101789      0.126847     Biso  1.000000 H
   H9         1.0     0.246311      0.565451      0.068522     Biso  1.000000 H
   H10        1.0     0.262572      0.452393      0.892769     Biso  1.000000 H
   H11        1.0     0.755562      0.972591      0.987199     Biso  1.000000 H
   H12        1.0     0.793350      0.890157      0.806544     Biso  1.000000 H
   H13        1.0     0.403447      0.264701      0.750488     Biso  1.000000 H
   H14        1.0     0.620422      0.041172      0.687759     Biso  1.000000 H
   H15        1.0     0.092556      0.520808      0.668540     Biso  1.000000 H
   H16        1.0     0.878938      0.425531      0.782733     Biso  1.000000 H
   H17        1.0     0.277244      0.787151      0.721416     Biso  1.000000 H
   H18        1.0     0.271302      0.927730      0.894763     Biso  1.000000 H
   H19        1.0     0.759342      0.441358      0.956243     Biso  1.000000 H
   H20        1.0     0.752188      0.576468      0.127365     Biso  1.000000 H
   H21        1.0     0.268047      0.035282      0.037284     Biso  1.000000 H
   H22        1.0     0.275885      0.149754      0.210990     Biso  1.000000 H
   H23        1.0     0.746530      0.683086      0.267856     Biso  1.000000 H
   H24        1.0     0.761337      0.795423      0.443837     Biso  1.000000 H
   H25        1.0     0.255738      0.275725      0.350865     Biso  1.000000 H
   H26        1.0     0.293096      0.355889      0.531770     Biso  1.000000 H
   H27        1.0     0.900278      0.982765      0.586210     Biso  1.000000 H
   H28        1.0     0.117928      0.205148      0.649812     Biso  1.000000 H
   K1         1.0     0.521894      0.844904      0.133478     Biso  1.000000 K
   K2         1.0     0.022109      0.404647      0.202560     Biso  1.000000 K
   K3         1.0     0.050181      0.620362      0.494791     Biso  1.000000 K
   K4         1.0     0.525553      0.061692      0.424774     Biso  1.000000 K
   K5         1.0     0.551396      0.626266      0.841003     Biso  1.000000 K
   K6         1.0     0.025080      0.187929      0.910496     Biso  1.000000 K
\end{verbatim}
\end{tiny}

\subsubsection{Structure \textbf{D}}
\begin{tiny}
\begin{verbatim}
#======================================================================

# CRYSTAL DATA

#----------------------------------------------------------------------

STRUCTURE_D


_pd_phase_name                         'EA425   7.712  7.595 14.392 75.284 91.'
_cell_length_a                         7.72022
_cell_length_b                         7.60670
_cell_length_c                         14.42992
_cell_angle_alpha                      75.33833
_cell_angle_beta                       91.81866
_cell_angle_gamma                      90.08340
_symmetry_space_group_name_H-M         'P 1'
_symmetry_Int_Tables_number            1

loop_
_symmetry_equiv_pos_as_xyz
   'x, y, z'

loop_
   _atom_site_label
   _atom_site_occupancy
   _atom_site_fract_x
   _atom_site_fract_y
   _atom_site_fract_z
   _atom_site_adp_type
   _atom_site_B_iso_or_equiv
   _atom_site_type_symbol
   C1         1.0     0.353113      0.790504      0.595693     Biso  1.000000 C
   C2         1.0     0.242176      0.862362      0.653863     Biso  1.000000 C
   C3         1.0     0.351241      0.855667      0.493509     Biso  1.000000 C
   C4         1.0     0.125018      0.010172      0.606357     Biso  1.000000 C
   C5         1.0     0.244709      0.797781      0.758797     Biso  1.000000 C
   C6         1.0     0.126201      0.073727      0.503548     Biso  1.000000 C
   C7         1.0     0.018434      0.088759      0.663009     Biso  1.000000 C
   C8         1.0     0.238373      0.997420      0.448767     Biso  1.000000 C
   C9         1.0     0.131091      0.883361      0.814426     Biso  1.000000 C
   C10        1.0     0.022349      0.025094      0.764836     Biso  1.000000 C
   C11        1.0     0.138295      0.821532      0.920882     Biso  1.000000 C
   C12        1.0     0.359086      0.659200      0.809370     Biso  1.000000 C
   C13        1.0     0.367373      0.601824      0.910994     Biso  1.000000 C
   C14        1.0     0.259199      0.676370      0.968693     Biso  1.000000 C
   C15        1.0     0.270318      0.617560      0.073846     Biso  1.000000 C
   C16        1.0     0.036410      0.901415      0.977752     Biso  1.000000 C
   C17        1.0     0.161137      0.706511      0.129143     Biso  1.000000 C
   C18        1.0     0.046339      0.845606      0.079742     Biso  1.000000 C
   C19        1.0     0.178141      0.653156      0.232227     Biso  1.000000 C
   C20        1.0     0.387481      0.480870      0.124898     Biso  1.000000 C
   C21        1.0     0.402905      0.431332      0.227498     Biso  1.000000 C
   C22        1.0     0.298385      0.518718      0.279595     Biso  1.000000 C
   C23        1.0     0.889658      0.284360      0.132089     Biso  1.000000 C
   C24        1.0     0.770607      0.212024      0.074831     Biso  1.000000 C
   C25        1.0     0.902176      0.219273      0.234107     Biso  1.000000 C
   C26        1.0     0.659945      0.063601      0.122897     Biso  1.000000 C
   C27        1.0     0.759736      0.276127      0.969960     Biso  1.000000 C
   C28        1.0     0.675176      0.000403      0.225768     Biso  1.000000 C
   C29        1.0     0.546751      0.983814      0.066635     Biso  1.000000 C
   C30        1.0     0.795342      0.077107      0.279710     Biso  1.000000 C
   C31        1.0     0.638878      0.190392      0.914757     Biso  1.000000 C
   C32        1.0     0.536723      0.047966      0.964774     Biso  1.000000 C
   C33        1.0     0.632703      0.252795      0.808286     Biso  1.000000 C
   C34        1.0     0.868668      0.414471      0.919472     Biso  1.000000 C
   C35        1.0     0.864505      0.471856      0.818036     Biso  1.000000 C
   C36        1.0     0.748517      0.398024      0.760635     Biso  1.000000 C
   C37        1.0     0.747481      0.457777      0.655663     Biso  1.000000 C
   C38        1.0     0.523908      0.173647      0.751072     Biso  1.000000 C
   C39        1.0     0.631920      0.369576      0.599926     Biso  1.000000 C
   C40        1.0     0.521947      0.230463      0.649033     Biso  1.000000 C
   C41        1.0     0.637983      0.424445      0.496979     Biso  1.000000 C
   C42        1.0     0.859508      0.594661      0.605447     Biso  1.000000 C
   C43        1.0     0.862859      0.646710      0.502971     Biso  1.000000 C
   C44        1.0     0.751594      0.561165      0.450200     Biso  1.000000 C
   H1         1.0     0.445516      0.682716      0.629315     Biso  1.000000 H
   H2         1.0     0.437818      0.794814      0.451441     Biso  1.000000 H
   H3         1.0     0.037865      0.184510      0.467045     Biso  1.000000 H
   H4         1.0     0.932881      0.202920      0.629462     Biso  1.000000 H
   H5         1.0     0.236196      0.051794      0.371194     Biso  1.000000 H
   H6         1.0     0.937351      0.095527      0.802828     Biso  1.000000 H
   H7         1.0     0.447386      0.593609      0.770523     Biso  1.000000 H
   H8         1.0     0.462336      0.496481      0.944239     Biso  1.000000 H
   H9         1.0     0.946028      0.012004      0.945935     Biso  1.000000 H
   H10        1.0     0.965577      0.916357      0.119821     Biso  1.000000 H
   H11        1.0     0.096415      0.719953      0.274310     Biso  1.000000 H
   H12        1.0     0.472311      0.412639      0.085577     Biso  1.000000 H
   H13        1.0     0.496714      0.327119      0.264242     Biso  1.000000 H
   H14        1.0     0.311010      0.483002      0.357703     Biso  1.000000 H
   H15        1.0     0.978079      0.391881      0.097841     Biso  1.000000 H
   H16        1.0     0.994978      0.280479      0.275341     Biso  1.000000 H
   H17        1.0     0.592462      0.888688      0.262850     Biso  1.000000 H
   H18        1.0     0.466066      0.868750      0.100445     Biso  1.000000 H
   H19        1.0     0.808257      0.021449      0.357135     Biso  1.000000 H
   H20        1.0     0.446578      0.977279      0.927117     Biso  1.000000 H
   H21        1.0     0.962599      0.479151      0.958208     Biso  1.000000 H
   H22        1.0     0.956150      0.576710      0.784755     Biso  1.000000 H
   H23        1.0     0.437143      0.062617      0.782587     Biso  1.000000 H
   H24        1.0     0.436453      0.159770      0.608728     Biso  1.000000 H
   H25        1.0     0.551701      0.358726      0.454268     Biso  1.000000 H
   H26        1.0     0.949486      0.661358      0.645320     Biso  1.000000 H
   H27        1.0     0.952580      0.751537      0.467078     Biso  1.000000 H
   H28        1.0     0.752143      0.601459      0.372029     Biso  1.000000 H
   K1         1.0     0.209972      0.447443      0.552994     Biso  1.000000 K
   K2         1.0     0.256789      0.084521      0.179527     Biso  1.000000 K
   K3         1.0     0.712376      0.988846      0.547488     Biso  1.000000 K
   K4         1.0     0.749766      0.623699      0.173753     Biso  1.000000 K
   K5         1.0     0.217819      0.267693      0.868463     Biso  1.000000 K
   K6         1.0     0.716457      0.804007      0.859901     Biso  1.000000 K
\end{verbatim}
\end{tiny}

\subsubsection{Structure \textbf{PS--K$_3$Picene}}
\begin{tiny}
\begin{verbatim}
#======================================================================

# CRYSTAL DATA

#----------------------------------------------------------------------

STRUCTURE_PS_K3PC


_pd_phase_name                         'Herringbon                            '
_cell_length_a                         7.65312
_cell_length_b                         7.70746
_cell_length_c                         14.35388
_cell_angle_alpha                      90
_cell_angle_beta                       106.19752
_cell_angle_gamma                      90
_symmetry_space_group_name_H-M         'P 1'
_symmetry_Int_Tables_number            1

loop_
_symmetry_equiv_pos_as_xyz
   'x, y, z'

loop_
   _atom_site_label
   _atom_site_occupancy
   _atom_site_fract_x
   _atom_site_fract_y
   _atom_site_fract_z
   _atom_site_adp_type
   _atom_site_B_iso_or_equiv
   _atom_site_type_symbol
   C1         1.0     0.567357      0.257521      0.280362     Biso  1.000000 C
   C2         1.0     0.432643      0.757521      0.719638     Biso  1.000000 C
   C3         1.0     0.427028      0.144468      0.229183     Biso  1.000000 C
   C4         1.0     0.572972      0.644468      0.770817     Biso  1.000000 C
   C5         1.0     0.375462      0.130417      0.125692     Biso  1.000000 C
   C6         1.0     0.624538      0.630417      0.874308     Biso  1.000000 C
   C7         1.0     0.463733      0.233342      0.072753     Biso  1.000000 C
   C8         1.0     0.536267      0.733342      0.927247     Biso  1.000000 C
   C9         1.0     0.601271      0.350172      0.120170     Biso  1.000000 C
   C10        1.0     0.398729      0.850172      0.879830     Biso  1.000000 C
   C11        1.0     0.656389      0.366014      0.224391     Biso  1.000000 C
   C12        1.0     0.343611      0.866014      0.775609     Biso  1.000000 C
   C13        1.0     0.796463      0.479219      0.274527     Biso  1.000000 C
   C14        1.0     0.203537      0.979219      0.725473     Biso  1.000000 C
   C15        1.0     0.853035      0.487690      0.377907     Biso  1.000000 C
   C16        1.0     0.146965      0.987690      0.622093     Biso  1.000000 C
   C17        1.0     0.775786      0.382737      0.435551     Biso  1.000000 C
   C18        1.0     0.224214      0.882737      0.564449     Biso  1.000000 C
   C19        1.0     0.629348      0.264413      0.386719     Biso  1.000000 C
   C20        1.0     0.370652      0.764413      0.613281     Biso  1.000000 C
   C21        1.0     0.555026      0.153895      0.444207     Biso  1.000000 C
   C22        1.0     0.444974      0.653895      0.555793     Biso  1.000000 C
   C23        1.0     0.615529      0.155769      0.546912     Biso  1.000000 C
   C24        1.0     0.384471      0.655769      0.453088     Biso  1.000000 C
   C25        1.0     0.757597      0.264841      0.598741     Biso  1.000000 C
   C26        1.0     0.242403      0.764841      0.401259     Biso  1.000000 C
   C27        1.0     0.841834      0.382344      0.543474     Biso  1.000000 C
   C28        1.0     0.158166      0.882344      0.456526     Biso  1.000000 C
   C29        1.0     0.986056      0.486363      0.595128     Biso  1.000000 C
   C30        1.0     0.013944      0.986363      0.404872     Biso  1.000000 C
   C31        1.0     0.053102      0.481779      0.698469     Biso  1.000000 C
   C32        1.0     0.946898      0.981779      0.301531     Biso  1.000000 C
   C33        1.0     0.976641      0.369887      0.754639     Biso  1.000000 C
   C34        1.0     0.023359      0.869887      0.245361     Biso  1.000000 C
   C35        1.0     0.826556      0.257333      0.705006     Biso  1.000000 C
   C36        1.0     0.173444      0.757333      0.294994     Biso  1.000000 C
   C37        1.0     0.756822      0.140058      0.762021     Biso  1.000000 C
   C38        1.0     0.243178      0.640058      0.237979     Biso  1.000000 C
   C39        1.0     0.826558      0.130877      0.865296     Biso  1.000000 C
   C40        1.0     0.173442      0.630877      0.134705     Biso  1.000000 C
   C41        1.0     0.970398      0.239307      0.912300     Biso  1.000000 C
   C42        1.0     0.029602      0.739307      0.087700     Biso  1.000000 C
   C43        1.0     0.044220      0.358185      0.858701     Biso  1.000000 C
   C44        1.0     0.955780      0.858185      0.141299     Biso  1.000000 C
   H1         1.0     0.357794      0.061301      0.269270     Biso  1.000000 H
   H2         1.0     0.642206      0.561301      0.730730     Biso  1.000000 H
   H3         1.0     0.269535      0.038629      0.089086     Biso  1.000000 H
   H4         1.0     0.730465      0.538629      0.910914     Biso  1.000000 H
   H5         1.0     0.424606      0.222663      0.993859     Biso  1.000000 H
   H6         1.0     0.575394      0.722663      0.006141     Biso  1.000000 H
   H7         1.0     0.666096      0.432207      0.077026     Biso  1.000000 H
   H8         1.0     0.333904      0.932207      0.922974     Biso  1.000000 H
   H9         1.0     0.865762      0.560989      0.233704     Biso  1.000000 H
   H10        1.0     0.134238      0.060989      0.766296     Biso  1.000000 H
   H11        1.0     0.963246      0.578008      0.410145     Biso  1.000000 H
   H12        1.0     0.036754      0.078008      0.589855     Biso  1.000000 H
   H13        1.0     0.448163      0.061501      0.410103     Biso  1.000000 H
   H14        1.0     0.551837      0.561501      0.589897     Biso  1.000000 H
   H15        1.0     0.550427      0.065562      0.585538     Biso  1.000000 H
   H16        1.0     0.449573      0.565562      0.414462     Biso  1.000000 H
   H17        1.0     0.055437      0.574880      0.557872     Biso  1.000000 H
   H18        1.0     0.944563      0.074880      0.442128     Biso  1.000000 H
   H19        1.0     0.169008      0.564218      0.733373     Biso  1.000000 H
   H20        1.0     0.830992      0.064218      0.266627     Biso  1.000000 H
   H21        1.0     0.647428      0.050863      0.726742     Biso  1.000000 H
   H22        1.0     0.352572      0.550863      0.273258     Biso  1.000000 H
   H23        1.0     0.767833      0.039149      0.906405     Biso  1.000000 H
   H24        1.0     0.232167      0.539149      0.093595     Biso  1.000000 H
   H25        1.0     0.031640      0.228779      0.990349     Biso  1.000000 H
   H26        1.0     0.968360      0.728779      0.009651     Biso  1.000000 H
   H27        1.0     0.158331      0.440872      0.897286     Biso  1.000000 H
   H28        1.0     0.841669      0.940872      0.102714     Biso  1.000000 H
   K1         1.0     0.228169      0.306031      0.491218     Biso  1.000000 K
   K2         1.0     0.771831      0.806031      0.508782     Biso  1.000000 K
   K3         1.0     0.422623      0.289761      0.805023     Biso  1.000000 K
   K4         1.0     0.577377      0.789761      0.194977     Biso  1.000000 K
   K5         1.0     0.031107      0.273438      0.177182     Biso  1.000000 K
   K6         1.0     0.968893      0.773438      0.822818     Biso  1.000000 K
\end{verbatim}
\end{tiny}

\end{document}